\documentclass[MSc,science,english]{sscunictthesis} 
\usepackage{lipsum}
\usepackage[Lenny]{fncychap}
\usepackage[absolute,overlay]{textpos} % only used by lastpage.tex
\usepackage[force]{feynmp-auto}
\usepackage{float}
\usepackage{bbm}
\usepackage{amsfonts}
\usepackage{amssymb}
\usepackage{physics}
\usepackage{slashed}
\usepackage{wasysym}
\usepackage{ulem}
\usepackage{graphicx}
\usepackage{subcaption}

\begin{document}

\author{Gabriele Parisi}
\title{Hot QCD: transport coefficients in strong and weak coupling regimes}
\aayear{2023/2024}

\begin{supervisors} % mandatory
   \supervisor{Chiar.mo}{Prof.}{Vincenzo Greco}
   \supervisor{}{Prof.}{Salvatore Plumari}
\end{supervisors}

%\begin{cosupervisors} % optional
%   \cosupervisor{Chiar.ma}{Prof.ssa}{E. Noether}
%   \cosupervisor{}{Dr.}{S. Kovalevskaya}
%\end{cosupervisors}

\begin{opponents} %mandatory
%   \opponent{Chiar.mo}{Prof.}{M. Planck}
   \opponent{}{Prof.}{Jörg Aichelin}
\end{opponents}

\maketitlepage

\hypersetup{%
   pdfsubject={Scuola Superiore di Catania,MSc thesis},
   pdfkeywords={test1,test2,test3}
}

\chapter*{\iflanguage{italian}{Riassunto}{Abstract}}
\addcontentsline{toc}{chapter}{\iflanguage{italian}{Riassunto}{Abstract}}

We evaluate the transport coefficients, namely the shear viscosity $\eta$ and the spatial diffusion coefficient of heavy quarks $D_s$, within a Quasi-Particle Model (QPM). In particular, for the first time this has been done by exploiting novel lQCD data at $N_f=3+1$. In such a framework, in which the gluons are massive, the interactions among quarks and gluons have been evaluated via the scattering matrices evaluated from tree level Feynman diagrams. As far as the evaluation of $D_s$ is concerned, the Fokker-Planck approximation for the relativistic Boltzmann equation has been employed for the diffusion of both charm and bottom quarks in either a $N_f=3$ and a $N_f=3+1$ bulk. We have developed a dual-component Chapman-Enskog formalism for the ratio $\eta/s$ (being $s$ the entropy density) of a mixture of quarks and gluons. The temperature dependence of the shear viscosity is then compared to the results from a Green-Kubo approach, based on the numerical resolution of the relativistic Boltzmann equation with a stochastic implementation of the collision integral. Finally, using the above results, we compare the strong and weak coupling regimes for the ratio $(2\pi T D_s)/(4\pi \eta/s)$.

\tableofcontents

\chapter*{\iflanguage{italian}{Introduzione}{Introduction}}
\addcontentsline{toc}{chapter}{\iflanguage{italian}{Introduzione}{Introduction}}

Quarks and gluons, which represent the elementary constituents within the theory of \textit{Quantum ChromoDynamics} (QCD), have been studied by means of deep-inelastic $e$-$p$ scattering collisions, via high energy heavy-ion and $p$-$\Bar{p}$ collisions, and also through the associate production of jets in $e^+$-$e^-$ annihilation processes for energies in the CM of a few TeV. For high values of energy ($\sqrt{s_{NN}}>100$ GeV) and transferred momentum ($p_T>20$ GeV), the property of asymptotic freedom of QCD, i.e. the fact that the coupling constant assumes small values at high energies, allows for a perturbative treatment of the interaction among partons. On the other hand, at energies not much higher than the QCD scale $\Lambda_{\text{QCD}}\sim 250$ MeV, the perturbative approach is not feasible: indeed, low energy QCD is still one of the open sectors in the Standard Model, in which many fundamental processes (like hadronization and the transition to either a deconfined state or a colour superconducting phase) occur. In particular, the QCD phase diagram is currently the subject of intense study which aims to clarify the peculiarities of the phase transition between ordinary matter and a deconfined matter of quarks and gluons: such state of matter is called \textit{Quark-Gluon Plasma} (QGP). The comprehension of such phenomenon would allow us to obtain information concerning the hadronization mechanism, which as of today does not have a satisfactory description yet, and also the expansion of the universe within the first 10-20 $\mu $s after the Big Bang. Moreover, such plasma state could also exist in very dense cosmological objects (like neutron stars), therefore we can assert that the study of such transition it is of both theoretical and experimental interest.\\
From the experimental point of view the only way to study the QGP phase is through heavy ion collisions occurring at ultra-relativistic energies: by varying the energy of the beams we can test different values of temperature and baryonic potential, therefore probing different regions of the QCD phase diagram. Such studies have been performed for decades (and still are) in the Large Hadron Collider (LHC) at CERN and in the Relativistic Heavy Ion Collider (RHIC) at BNL: indication of the existence of QGP has been provided by such experiments in collisions of 200 AGeV or higher. All that being said, the very short lifetime of such deconfined state ($\sim 10$ fm/c  at LHC energies), as well as the impossibility to observe isolated quarks and gluons, do not allow for a direct analysis of the system. The only observables which are accessible are the hadrons detected in the final state, therefore the plasma state is ‘hidden' behind the hadronization process. Some quantities can however give us an indication on the formation of QGP and on its features, such quantities being the spectra of the detected particles and the collective flows. \\

From the theoretical point of view, the natural starting point for the study of such high-temperature region is of course perturbative QCD (pQCD). However, experimental evidence shows that in QGP the interaction among particles is stronger than expected, therefore the coupling constant is greater than the perturbative estimate at high energy. Due to this fact, approaches which take into account also non-perturbative dynamics have to be considered. One of such approaches is lattice QCD (lQCD), which allows for a numerical study of QCD in a space-time grid. This approach can however be used only for null baryon chemical potential: in the non-zero chemical potential region we have to consider effective models. All that being said, lQCD showed the presence of a crossover at $T_c\sim 155$ MeV$\sim 10^{12}$ K for a zero value of baryonic density. Phenomenological models show that for finite values of baryonic density such transition could occur at lower values of temperature, even if it is thought that such phase could show up as a \textit{superconductive colour phase}.\\
A good tool to study the properties of QGP is the propagation of heavy quarks (charm and bottom) in the plasma state. These particles, due to their high mass (1.3 GeV and 4.2 GeV, respectively) are produced in the very early stages after the collision, even before the formation of the QGP medium. Since these particles do not reach thermal equilibrium with the dense medium, the D and B mesons which are detected in the final state ‘keep memory' of the interaction of the heavy quarks with the QGP bulk. The starting point for the study of the charm and bottom dynamics in the QGP is the Boltzmann relativistic transport equation. In particular, the motion of the heavy quarks in the QGP, in the approximation of small momentum transfer, is usually treated as a Brownian motion and it is described by the \textit{Fokker-Planck equation}, which can be seen as an approximation of the Boltzmann equation. In this approach the effect of the interaction among the heavy quarks and the bulk is described in terms of the \textit{transport coefficients}, e.g. the \textit{spatial diffusion coefficient} $D_s$.\\
In the current research QGP can also be treated in terms of a hydrodynamic approach, in which a coefficient called \textit{shear viscosity} $\eta$ emerges as a proportionality factor for the shear viscous pressure tensor, encoding the non-ideal effects of interactions among two adjacent fluid sheets. Such coefficient may be derived by following several different approaches from the literature. One of them is called the \textit{Chapman-Enskog} (CE) approximation, in which we consider small deviations from the Boltzmann distribution at local equilibrium, and then expand in gradients with respect to such deviations: the formalism allows for various orders of approximation and also the treatment of non homogeneous mixtures. Another approach consists in a stochastic implementation of the collision integral of the relativistic Boltzmann equation, leading to an expression of $\eta$ in terms of the correlator of the shear momentum tensor at different times. The latter is known as the \textit{Green-Kubo} (GK) method.\\

To perform the evaluation of $D_s$ and $\eta$ we have to specify the nature of the bulk we are modeling. In order to go beyond the perturbative treatment for the QGP, which is believed to be a strongly coupled medium, this thesis work will deal with the \textit{Quasi-Particle Model} (QPM), an effective theory in which the strength of the interaction among the particles of the QGP bulk is encoded in their masses. In particular, the parton masses acquire a temperature dependence, then the coupling constant $g=g(T)$ is obtained via a fit to the lQCD thermodynamics data.\\
The relevant features of this work consist in the application of the two component Chapman-Enskog formalism within the QPM, where the interactions have been considered by evaluating the proper tree-level scattering matrices $\mathcal{M}$. In those we have to take into account the non-zero mass of the gluon, which of course affects the Feynman rules themselves. Then, our results have been compared to the ones from the Green-Kubo method, for which a  transport code has been developed.\\

This work will be organized as follows. In Chapter 1, the Quasi-Particle Model will be introduced, along with its predictions for the coupling constant, for the effective masses and for the ratio $P/\epsilon$. Moreover, we deal with the evaluation of the scattering matrix $\mathcal{M}$ within the QPM (for the details of the calculations we refer to Appendix A). In Chapter 2, we elaborate on the dynamics of heavy quarks within the QGP medium, therefore the Boltzmann transport equation and its approximation, the Fokker-Planck equation, will be described. Here we give a brief description of the spatial diffusion coefficient $D_s$. In Chapter 3, we introduce the Chapman-Enskog formalism for the shear viscosity $\eta$: after mentioning the role of this quantity within viscous hydrodynamics, we illustrate the CE formalism for both one and two component fluids. In Chapter 4, we show how the same quantity can be derived within the Green-Kubo formalism, by illustrating the details of such numerical procedure. Finally, in Chapter 5 after setting up the GK method (by applying to few simpler cases), we show the results obtained in the full physical case for $D_s$, for $\eta/s$ (being $s$ the entropy density) and for the ratio $(2\pi T D_s)/(4\pi \eta/s)$. The latter quantity has been conjectured to span between the values $1$ and $5/2$ while going from a strong coupling regime to a weakly coupled one. The conclusions of this thesis will then be summarized.

\chapter{\iflanguage{italian}{Stato dell'arte}{The Quasi-Particle Model}}

%\section{\iflanguage{italian}{Titolo della sezione}{Section title}}

%\subsection{\iflanguage{italian}{Titolo del paragrafo}{Subsection title}}

\section{Main features of QPM}
\label{Main features of QPM}

QCD at very high temperatures presents a phase which is called Quark-Gluon Plasma (QGP): experiments performed at facilities like the Large Hadron Collider (LHC) at CERN and the Relativistic Heavy Ion Collider (RHIC) at BNL are able to quantify the properties of such a deconfined phase of QCD. Together with lattice calculations on QCD thermodynamics, which are reaching unprecedented levels of accuracy, we are able to obtain crucial information in order to shed light on the properties of strongly interacting matter under extreme conditions.\\

From a theoretical point of view, the range of temperatures which is reached at RHIC and LHC is not large enough with respect to the $\Lambda_{\text{QCD}}$. This implies that the coupling $\alpha_s$ is not sufficiently small and we cannot neglect non-perturbative aspects of the interaction. The medium created in the collision can therefore be described as a many-body system which is heavily screened and affected by non-perturbative effects even at very high temperatures. As a proof of that, the experimental data for the observables which are related to the heavy quark dynamics (like $R_{AA}$ and $v_2$) are not well reproduced within a perturbative approach. In particular, pQCD calculations underestimate the energy loss and the elliptic flow of the heavy quarks. More in general, the perturbative calculations accurately describe the general features of QGP in the region of high $p_T$ (greater than at least 15-20 GeV), whereas they underestimate the entity of the interactions in the region of small $p_T$.\\

In a pQCD approach the interactions (among bulk partons and of the heavy quarks with the bulk partons) are studied via the Combridge matrices \cite{Combridge_1979}, in which the t-channel divergence is cured via the introduction of a Debye screening mass $m_D=\sqrt{4\pi\alpha_s}T$. All of this will result in a coupling constant whose temperature dependence is, at the 2 loop level \cite{Kaczmarek_2005}:
$$g^{-2}(T)=2\beta_0\log \left(\frac{2\pi T}{aT_c}\right)+\frac{\beta_1}{\beta_0}\log\left[2\log\left(\frac{2\pi T}{aT_c}\right)\right],$$
with:
$$\beta_0=\frac{1}{16\pi^2}\left(11-\frac23 N_f\right)~~~~~~\beta_1=\frac{1}{(16\pi^2)^2}\left(102-\frac{38}{3}N_f\right),$$
being $a=1.3$, $N_f$ the number of flavours and $T_c$ the critical temperature.\\
Such a treatment, however, does not take into account for non-perturbative aspects of the interaction. In the attempt of considering such aspects of the deconfined state, we have to get over the concept of massless and weakly interacting bulk partons. In \cite{Braaten_1991} it has been shown that in a QED medium with finite temperature the Born approximation is not appropriate for low momentum transfer: one solution is provided by the so-called \textit{hard thermal loop} (HTL) approach, which has then been extended to QCD \cite{Gossiaux_2008}. Another approach consists in describing the quarks and the gluons of the QGP bulk as an ensemble of \textit{quasi-particles} close to the critical temperature.\\

Indeed, one of the main concerns when dealing with the lattice results is interpreting them in terms of the appropriate degrees of freedom. In particular, a quite intriguing possibility is to bring the description of lattice results in terms of quark and gluon quasi-particles: if we encode most of the interaction in terms of effective masses depending on the temperature, we can treat the ‘residual' interaction among quasi-particles perturbatively. This will then allow for a description of the lattice QCD results of equilibrium thermodynamics, such as the Equation of State (EoS).\\
After performing a fit to the available lattice data, we will obtain the quark and gluon temperature-dependent masses (which are related to each other through the coupling constant) and a temperature-dependent \textit{bag pressure} constant, which is necessary for thermodynamic consistency and takes into account for further non-perturbative effects. Once this is done, we can extract the dynamical properties of our medium.\\

The temperature-dependent effective masses for quarks and gluons can be evaluated in a perturbative approach which, in absence of any chemical potential, suggests the relations \cite{PlumariQPM}:
\begin{equation}
m_g^2=\frac{1}{6}\left(N_c+\frac12 N_f\right)g^2T^2,~~~~~m_{q}^2=\frac{N_c^2-1}{8N_c}g^2T^2,
\label{3.9}
\end{equation}
where $N_f$ is the number of flavours considered and $N_c$ is the number of colours.
The coupling $g=g(T)$ is temperature-dependent, and the dependence cannot be determined via perturbation theory, but rather through a fit to lattice QCD data.\\
The pressure of the system can then be written as the sum of independent contributions coming from the different constituents (each having the $T$-dependent effective mass in \eqref{3.9}), plus a bag constant:
\begin{equation}
    P_{qp}(m_u,m_d,\dots,T)=\sum_{i=u,d,s,g}d_i\int \frac{\dd^3 \mathbf{p}}{(2\pi)^3}\frac{\mathbf{p}^2}{3E_i(p)}f_i(p)-B(T),
    \label{P_QPM}
\end{equation}
where $f_i(p)=[1\mp \exp[\beta E_i(p)]]^{-1}$ are the Bose/Fermi distribution functions for gluons and quarks, respectively, $E_i(p)=\sqrt{\mathbf{p}^2+m_i^2}$ and $d_i$ are the degeneracy factors, equal to $2\cdot 2\cdot N_c$ for quarks and to $2\cdot (N_C^2-1)$ for gluons. As previously mentioned, the bag term $B(T)$ has to be taken into account in order to have thermodynamic consistency, i.e. the following relation has to be satisfied:
$$\left(\frac{\partial P_{qp}}{\partial m_i}\right)_{T,\mu}=0,~~~i=u,d,\dots$$
which leads to a set of equations of the form:
\begin{equation}
\frac{\partial B}{\partial m_i}+d_i \int \frac{\dd^3 \mathbf{p}}{(2\pi)^3}\frac{m_i}{E_i(p)}f_i(p)=0.
\label{3.10}
\end{equation}
Notice that only one of the above equations is independent, since all the masses depend on the coupling $g$. The energy density of the system is then obtained from the pressure by making use of the thermodynamic relation $\epsilon(T)=TdP(T)/dT-P(T)$, leading to the expression:
\begin{align*}
    \epsilon_{qp}(T)&=\sum_id_i \int \frac{\dd^3 \mathbf{p}}{(2\pi)^3}E_i(p) f_i(p)+B(T)\equiv\\
    &\equiv\sum_i \epsilon_{\text{kin}}^i(m_i(T),T)+B(m_i(T)).
\end{align*}
Notice that the dependence on the temperature in all the above expressions can be both explicit (via $\beta\equiv T^{-1}$) and implicit (via the masses). The bag constant $B$ depends on $T$ only via $m_i$, since it is determined by equation \eqref{3.10}.\\

The above steps allow us to study the EoS in this \textit{Quasi-Particle Model} (QPM). We have expressed every quantity in terms of two unknown functions, namely $g(T)$ and $B(T)$, but we can see that actually even these functions are not independent one another, since \eqref{3.10} must hold. This means that only one function needs to be determined, and this is done by imposing the condition:
\begin{equation}
    \epsilon_{qp}(T)=\epsilon_{\text{lattice}}(T),
    \label{3.11}
\end{equation}
i.e. by performing a fit to the lattice data. By following the steps in \cite{Bozek_1998}, we can derive both sides of \eqref{3.11} with respect to $T$, recall equation \eqref{3.10} and use the expressions \eqref{3.9} for the QPM masses $m_i=m_i(T)$ in order to obtain:\footnote{Here (with little abuse of notation) $\partial m_i/\partial T$ refers to the derivative of $m_i$ with respect to the explicit dependence on $T$ only. For instance, for our QPM masses of the form $m_i(T)=\alpha_i g(T)T$, we consider $\partial m_i/\partial T=\alpha_i g(T)$.}
\begin{equation}
    \left.\frac{dg(T)}{dT}\right|_{\mu=0}=g\cdot \left\{\frac{\frac{d\epsilon_{\text{lattice}}}{dT}-\frac{\partial \epsilon_{\text{kin}}}{\partial T}-\sum_i\frac{\partial \epsilon_{\text{kin}}^i}{\partial m_i}\frac{\partial m_i}{\partial T}-\sum_i\frac{\partial B}{\partial m_i}\frac{\partial m_i}{\partial T}}{
    \sum_i \frac{\partial \epsilon_{\text{kin}}^i}{\partial m_i}m_i+\sum_i\frac{\partial B}{\partial m_i}m_i}\right\},
\label{3.12}
\end{equation}
where $\epsilon_{\text{kin}}=\sum_i \epsilon_{\text{kin}}^i$.
By performing the fit to the lattice energy density (for which the data from \cite{Borsanyi_2010} have been used,\footnote{Later on we will also perform a similar fit to the $N_f=3+1$ energy density lattice data.} $N_f=N_c=3$)\footnote{Usually in the literature for the strange quark one adds up the current physical mass, i.e. $m_s^2=\frac{N_c^2-1}{8N_c}g^2T^2+m_{0s}^2$. However, here we considered $m_s=m_{u,d}$, since the effect of a finite current mass for the strange quark is negligible for the description of thermodynamic quantities.} we can therefore solve the differential equation \eqref{3.12} by means of the Euler method, given the initial condition $g(T_0)=g_0$. For $T_0=151$ MeV and $g_0=6.26$ we get a function which can then be fitted as:\footnote{The initial value for $g_0$ has been set in order to match the lattice data for the pressure at temperature $T_0$.}
\begin{align}
    g(T)&=(c_0+c_1T+c_2T^2)\exp[-c_3\cdot T^{c_4}]~~~~~\text{for }T\leq 0.22 \text{ GeV},\nonumber\\
g(T)&=\sqrt{\frac{16\pi^2}{9\log[(b_0(T/b_1-b_2))^2]}}~~~~~~~~~~~~~\text{ for }T\geq 0.22 \text{ GeV},
\label{gfit}
\end{align}
\begin{table}[]
    \centering
    \begin{tabular}{|c|c|c|c|c|c|c|c|}
        \hline
         $c_0$&$c_1$&$c_2$&$c_3$&$c_4$&$b_0$&$b_1$&$b_2$  \\\hline
         $46.737$&$-436.86$&$1266.4$&$77.763$&$2.7523$&$2.2800$&$0.155$&$0.55$ \\\hline
    \end{tabular}
    \caption{Values for the constants used in the fit for $g(T)$.}
    \label{tab:1}
\end{table}
where the values of the constants are reported in Table \ref{tab:1}.\\

\begin{figure}[ht]
\begin{minipage}[ht]{0.49\textwidth}
\centering
    \includegraphics[scale=0.30]{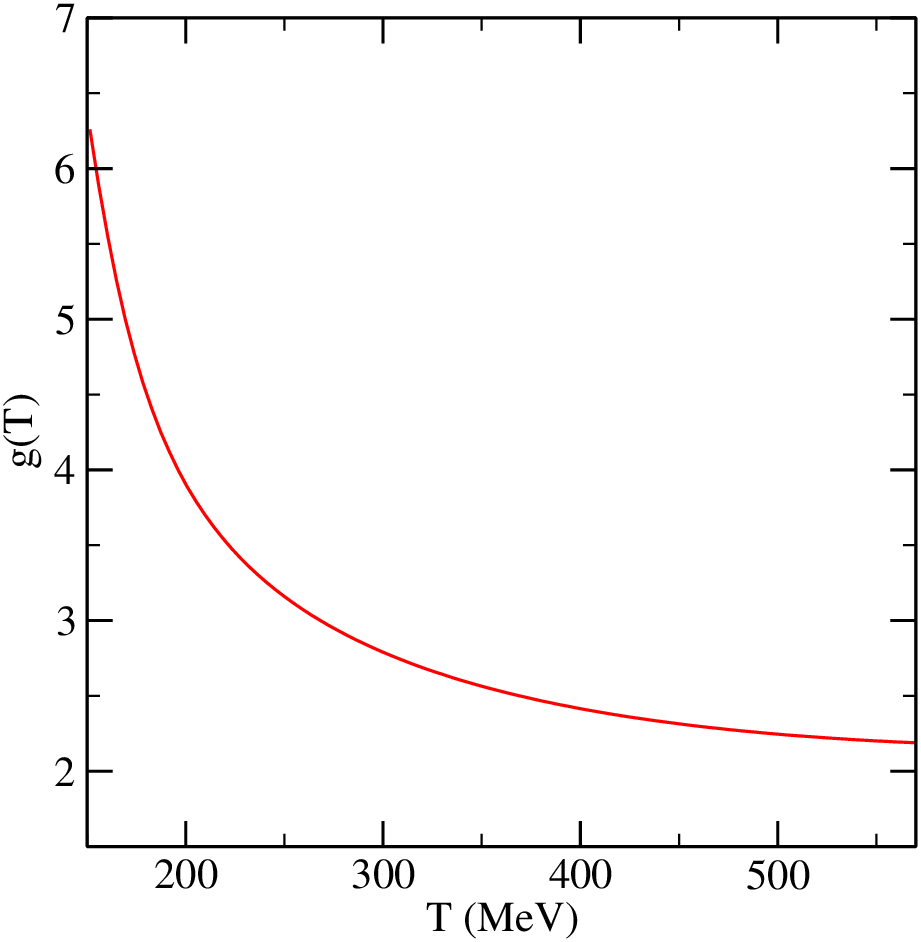}
\end{minipage}    
\hfil
\begin{minipage}[ht]{0.49\textwidth}
\centering
	\includegraphics[scale=0.30]{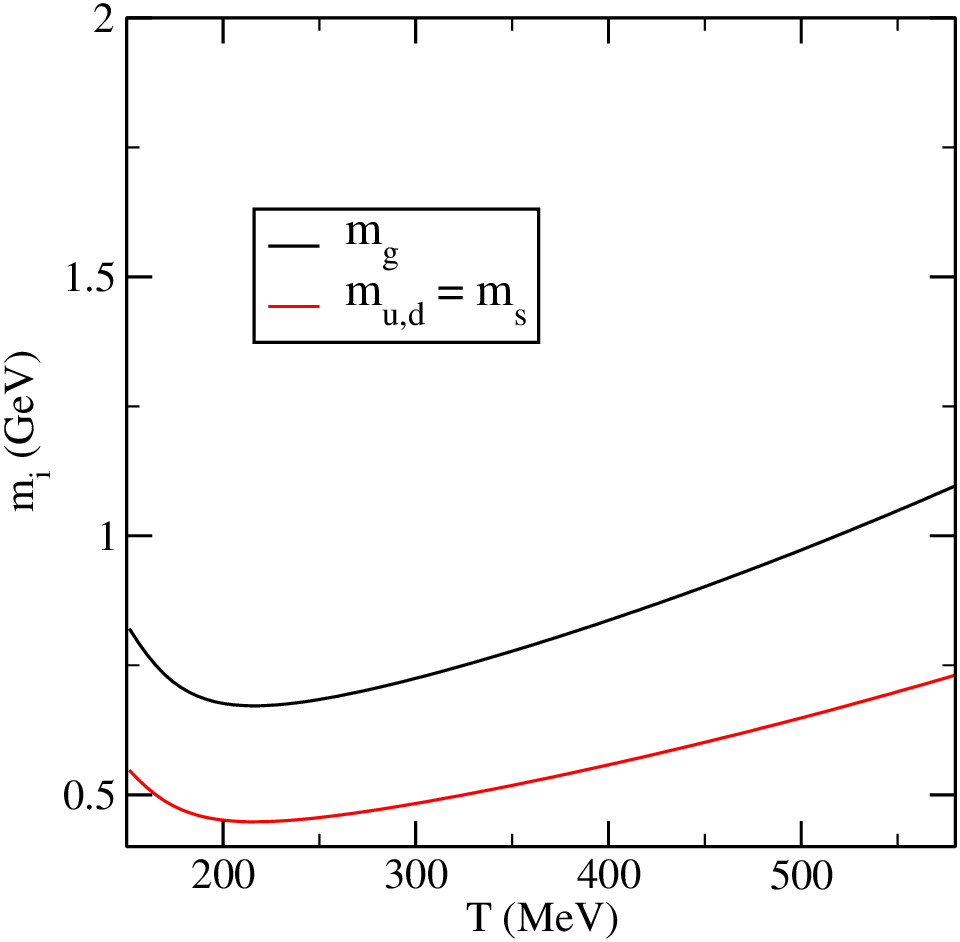}
\end{minipage} 
\caption{\textit{Left panel}: the coupling $g$ as a function of $T$. \textit{Right panel}: the gluon and quark Quasi-Particle masses as functions of $T$. Notice that we considered $m_s=m_{u,d}$, since the effect of a finite current mass for the strange quark is negligible for the description of thermodynamic quantities.}
    \label{fig1.1}
\end{figure}

In Figure \ref{fig1.1} we show the temperature dependence of the coupling $g$ and of the quark and gluon masses that we obtain from this procedure. We notice that at sufficiently high temperatures $m(T)\sim T$, as one can expect since in these regimes $T$ remains the only scale of the system.\\

\begin{figure}[ht]
    \centering
    \includegraphics[scale=0.4]{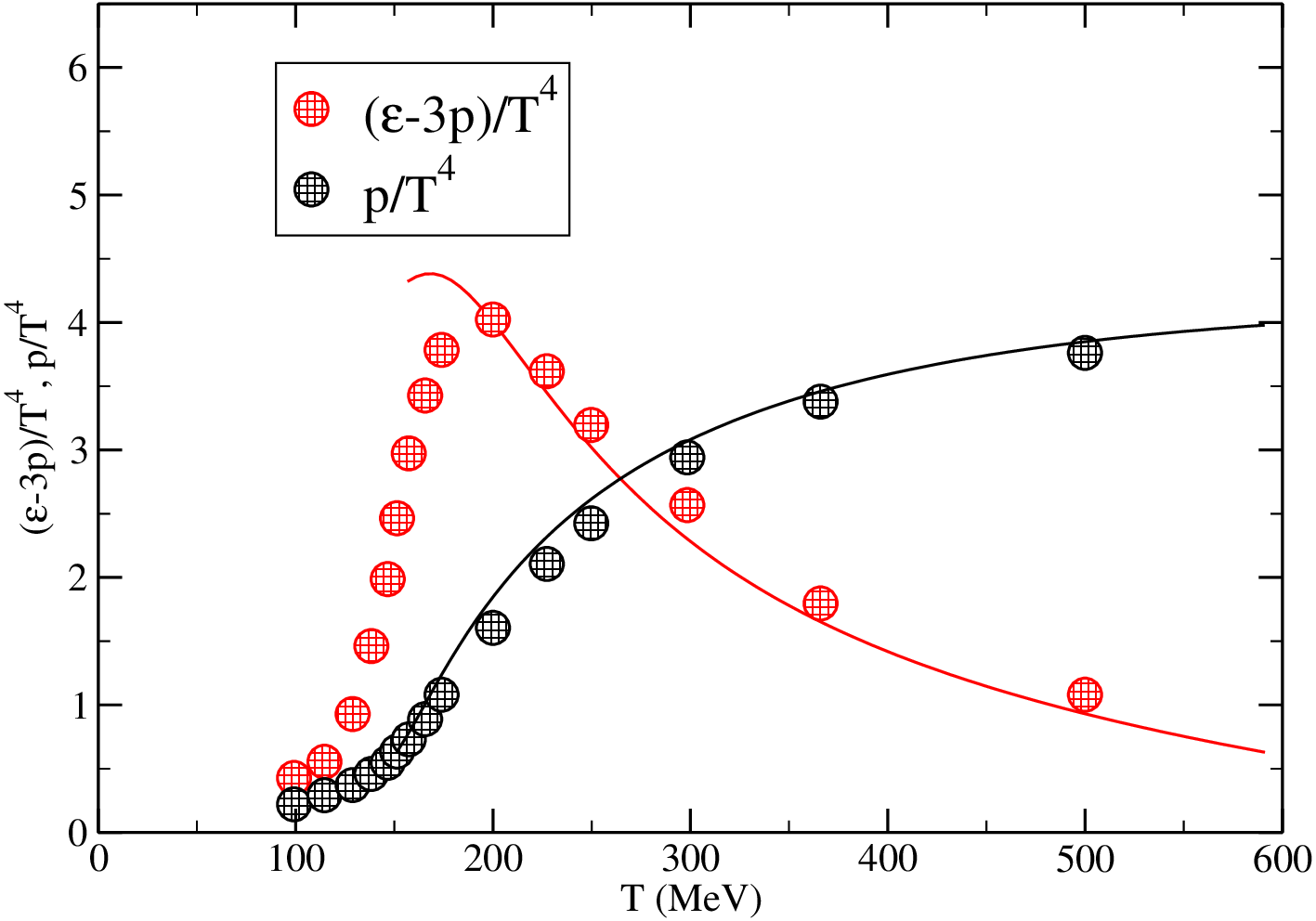}
    \caption{Lattice data for the normalized pressure and trace anomaly as functions of the temperature (dots), together with the Quasi-Particle Model curves. The lattice data are in the $N_f=3$ case, and are taken from the Wuppertal-Budapest collaboration \cite{Borsanyi_2010}.}
    \label{fig4.2}
\end{figure}

\begin{figure}[t]
\begin{minipage}[t]{0.49\textwidth}
    \includegraphics[scale=0.35]{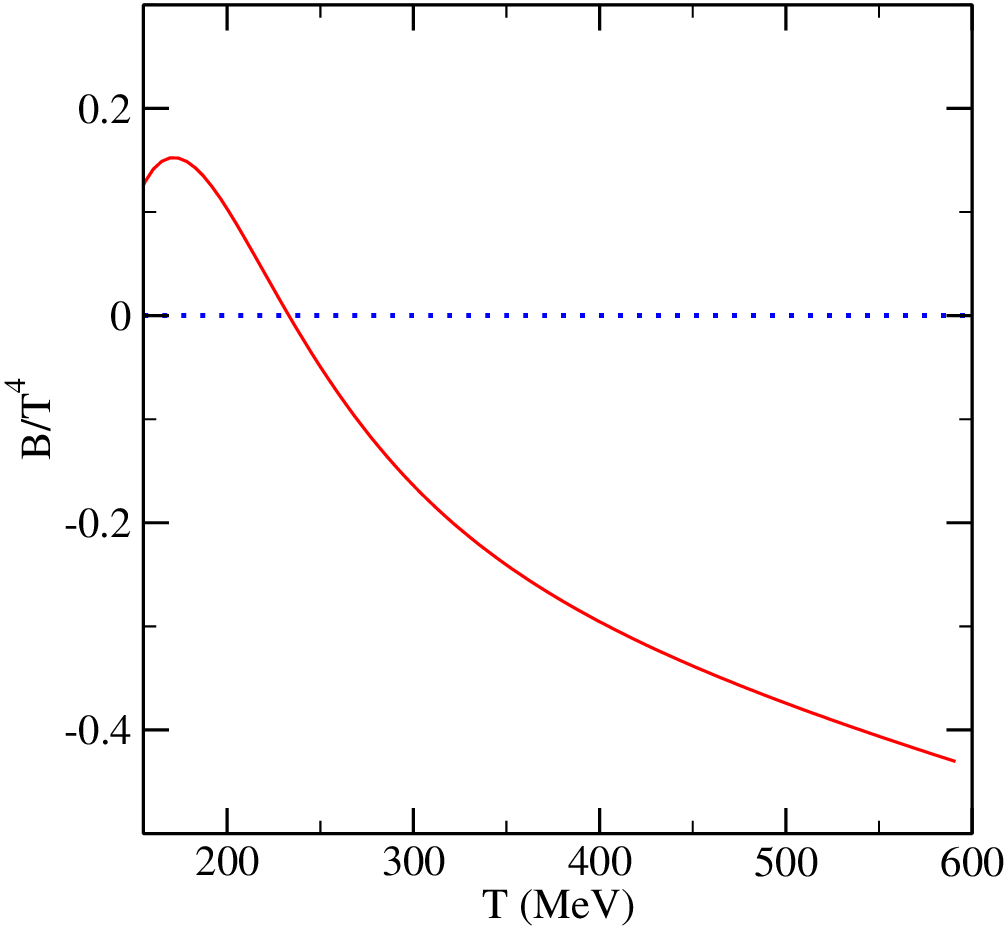}
\end{minipage}    
\hfil
\begin{minipage}[t]{0.49\textwidth}
	\includegraphics[scale=0.35]{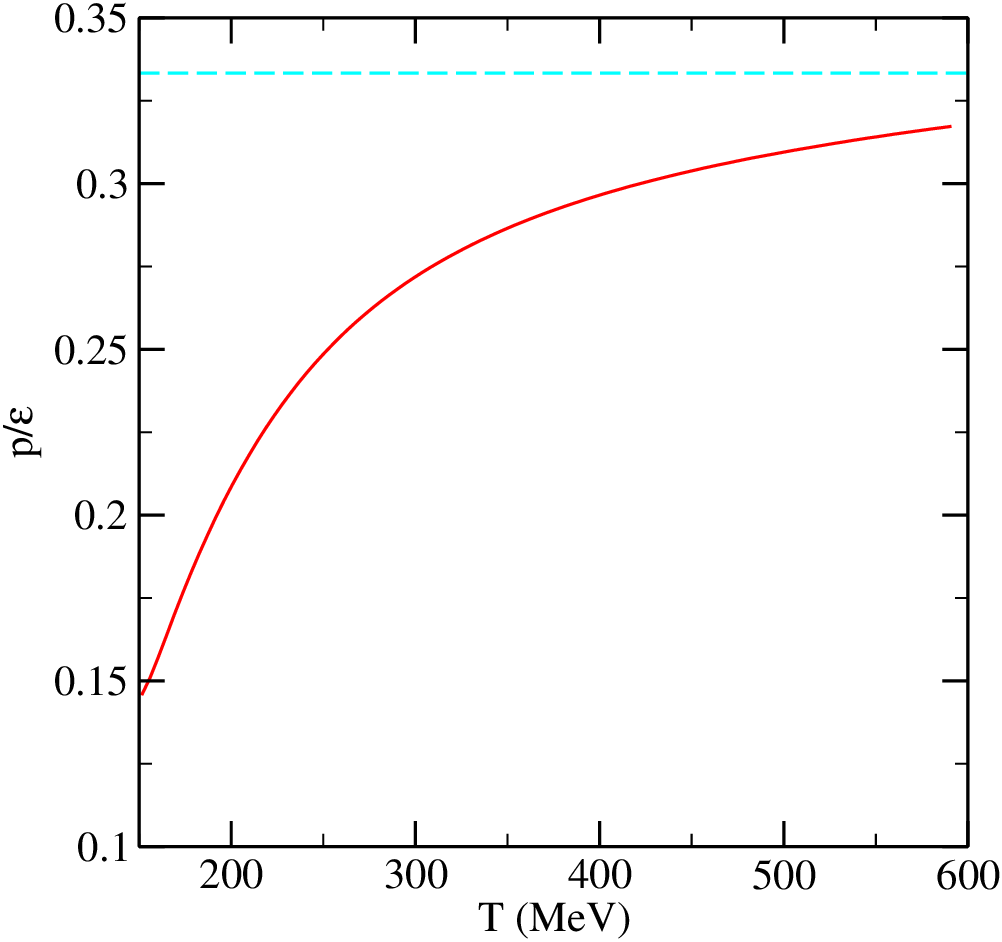}
\end{minipage}
\caption{\textit{Left panel}: the bag constant $B$ as a function of the temperature. \textit{Right panel}:
the ratio $p/\epsilon$ as a function of the temperature, the light blue dotted line corresponds to the ideal case, in which $\epsilon=3p$.}
    \label{fig4.3}
\end{figure}

Having performed a fit to the lattice data for the energy density, we can then compare the lattice results for pressure and trace anomaly with the curves obtained in the QPM. In Figure \ref{fig4.2} we show a good agreement between these data, which outlines thermodynamic consistency. Moreover, the function $B(T)$ can be obtained from the difference between the lattice energy density and the kinetic contribution from our model. In particular:
$$B(T)=\epsilon_{\text{lattice}}(T)-\sum_i \epsilon_{\text{kin}}^i(T).$$
As we mentioned, the bag constant provides another measure for non-perturbative physics which cannot be absorbed into the effective quark and gluon masses.
We show the temperature dependence of $B$ in Figure \ref{fig4.3}. For completeness, in Figure \ref{fig4.3} we also show the ratio between pressure and energy density as a function of the temperature, together with the ideal gas case in which $\epsilon=3p$.

\section{Perturbative scattering of massive partons}
\label{Perturbative scattering of massive partons}

In such a framework, both quarks and gluons are massive. This means that, in order to calculate the matrix element $\mathcal{M}$ corresponding to a scattering in QPM, the usual scalar propagator for the gluon (which is used in pQCD) has to be replaced with a massive vector propagator. What we will deal with is therefore \cite{Moreau_2019}:
\begin{fmffile}{propagators}
\begin{align}
    \parbox{35mm}{\begin{fmfgraph*}(75,50)
    \fmfleft{i1}
    \fmfright{i2}
    \fmf{gluon, width=6, label=$q$, label.side=top}{i1,i2}
    \fmflabel{$\mu, a$}{i1}
    \fmflabel{$\nu, b$}{i2}
\end{fmfgraph*}\hspace*{4mm}}&=-\delta_{ab}\frac{g^{\mu\nu}-q^\mu q^\nu/M_g^2}{q^2-M_g^2}
\label{1.6}\\
\parbox{35mm}{\begin{fmfgraph*}(75,50)
    \fmfleft{i1}
    \fmfright{i2}
    \fmf{fermion, width=6, label=$q$, label.side=top}{i1,i2}
    \fmflabel{$i$}{i1}
    \fmflabel{$j$}{i2}
\end{fmfgraph*}\hspace*{4mm}}&=-\delta_{ij}\frac{\slashed{q}+M_q}{q^2-M_q^2}
\label{1.7}
\end{align}
\end{fmffile}
where $q$ is the 4-momentum of the exchanged particle. The delta functions (over $a,b$ for the gluon, over $i,j$ for the quark in the above diagrams) ensure that the exchanged gluon/quark is connected with the other parts of the diagram carrying the same colour.\\

In this work we are going to consider gluon-gluon, gluon-quark and quark-quark scattering, since we will need to understand both heavy quarks interacting with a bulk of quasi-particles (as in Chapter \ref{Heavy quark dynamics in QGP}) and the bulk partons themselves interacting with one another (see Chapters \ref{Shear viscosity in the Chapman-Enskog formalism} and \ref{Shear viscosity from the Green-Kubo formalism}). This will be achieved by evaluating the proper scattering matrices $\mathcal{M}$ using pQCD Feynman rules at tree level, for each process under consideration, but taking into account the non-zero mass of the exchanged boson. Of course, when needed, proper exchange and annihilation terms (as well as all their interferences) have to be considered. The details of this procedure for each of the cases under consideration will be thoroughly explained when employed. The invariant matrix element (squared) $|\mathcal{M}|^2$
is then calculated by averaging over initial - and summing over final - spins and colours.\\

Notice that the above procedure is substantially different from standard pQCD with massless gluons as far as two main points are concerned: first of all, for the gluon one has to consider \eqref{1.6} instead of the typical massless propagator (and of course also the massive mass-shell relation $k^2=M_g^2$ has to be employed when developing the calculations). Moreover, while evaluating $|\mathcal{M}|^2$ also the sum over polarizations $\lambda$ of the gluons has to be modified as:
$$\sum_\lambda \epsilon_\mu^\lambda(k)\epsilon_\nu^\lambda(k)=-g_{\mu\nu}+\frac{k_\mu k_\nu}{k^2},$$
whereas in usual pQCD only the $-g_{\mu\nu}$ term in the RHS would be present.\\
The results of our interest for the explicit squared scattering matrices can be found in Appendix \ref{Appendix A}.\\

As a final remark, note that the $g(T)$ appearing in the QPM masses \eqref{3.9} is in principle not equal to the coupling arising from the Feynman diagram vertices. Indeed, in order to reproduce the experimental data for the \textit{nuclear modification factor} $R_{AA}$ and for the \textit{elliptic flows} $v_n$ (e.g. $v_2$ and $v_3$), one finds that both coupling constants can be assumed equal to the (previously derived) $g(T)$, apart from a multiplicative correction $\sim$25\% to be applied to the coupling constant from the vertices. Operatively, this translates into a corrective factor $\propto g(T)^{-4}\sim 1/2.4$ to be applied to the observables which we will deal with, i.e. the spatial diffusion coefficient and the shear viscosity to entropy density ratio. Of course, when performing ratios among these quantities such a corrective factor cancels out.

\chapter{Heavy quark dynamics in QGP}
\label{Heavy quark dynamics in QGP}
In the context of QGP, heavy quarks and their bound states are considered very useful probes to study the evolution of such deconfined state. Their main feature is an high value of mass, which is much greater than the hadronization scale $\Lambda_{\text{QCD}}$: since in such regimes the coupling constant is sufficiently small, we are entitled to use a perturbative approach in the calculation of the cross sections.\\
More specifically, the quarks which serve our purpose are the \textit{charm}, whose bare mass is about 1.3 GeV, and the \textit{bottom}, which has a bare mass of 4.2 GeV \cite{Workman_2022}. The \textit{top} quark is usually not taken into account, since its huge mass (about 173 GeV) implies decay times smaller than 1 fm, therefore this particle is not even able to hadronize. On the other hand, the light quarks along with the gluons are the constituents of the expanding bulk.\\

The masses of the heavy quarks is not only greater than $\Lambda_{\text{QCD}}$, but also greater than the typical temperatures that characterize the system during the QGP formation. First of all, this implies that the number of heavy quark-antiquark pairs is determined in the first instants after the collision: indeed, the temperature is not high enough to determine a significant production of these pairs via thermal excitation of the vacuum.\\
Moreover, the thermalization time of the heavy quarks is much greater than the one of the light quarks and it is comparable with the lifetime of the QGP itself ($\tau_\text{QGP}\sim 10$ fm/c for typical Pb-Pb collisions at LHC energies \cite{Markert_2008}). This means that the charm (and even more, the bottom) are not able to reach local thermal equilibrium with the bulk. However, their interactions with the medium are able to modify the initial momentum spectra, so that the final hadrons give an indication of such interactions.\\
Finally, the \textit{thermal momentum} is greater than the momentum typically transferred by the medium, since we have:
$$p_{\text{thermal}}^2\sim 3m_QT\gg T^2$$
in a non-relativistic approximation. This allows us to study the evolution of heavy quarks in terms of a Brownian-like approach, more specifically we can expand the Boltzmann equation in order to obtain the Fokker-Planck equation, from which in turn we can study the drag and diffusion coefficients of the medium. Several studies have however shown that the Brownian motion approximation is not strongly valid in the range of momenta of interest for heavy
quarks $(p \approx 2-3 m_Q)$ and that, especially for the charm, the description of Brownian motion is in general a strong approximation, far
from the realistic system.

\section{The Boltzmann transport equation}
\label{The Boltzmann transport equation}
The starting point for heavy quark dynamics in QGP is of course analogous to the one we have in the kinetic theory of gases. Indeed, a purely deterministic treatment of a system with a huge number of elements would require the microscopic study of the exact equations of motion for each particle. For very large systems ($\sim10^{23}$ particles) such approach would be impossible, therefore we rather make use of a description in terms of distribution functions and probability density, which takes into account only for the macroscopic features of the system and allows for a probabilistic study of its evolution.\\

More specifically, the equations we are going to deal with are able to describe the time evolution of a system far from thermal equilibrium via a \textit{distribution function} in the \textit{phase space}, therefore as a function depending on both position and momentum. Such approach is well suited for the description of a Brownian-like motion.\\
Our starting point is therefore going to be the \textit{relativistic Boltzmann equation} for the distribution function $f(\mathbf{r},\mathbf{p},t)$:
\begin{equation}
    \frac{\partial f}{\partial t}+\frac{\mathbf{p}}{E}\cdot\nabla_\mathbf{r}f+\mathbf{F}^{\text{ext}}\cdot\nabla_\mathbf{p}f=\left(\frac{\partial f}{\partial t}\right)_{\text{coll}},
    \label{3.2}
\end{equation}
where the term in the RHS, called \textit{collision integral}, expresses the variations of the distribution function which are related to the collisions among particles.\\
At this point, in order to go further we have to consider some hypotheses. In particular, we assume that:
\begin{itemize}
    \item The internal forces are the predominant terms and the external forces are negligible, which is indeed the case of QGP.
    \item If the particles interact via a short range force, the probability of occurrence of collisions involving more than two particles is negligible with respect to the probability of binary collisions.
    \item The probability to find two particles with momenta $\mathbf{p}$ and $\mathbf{q}$ is the product of the single densities:
    $$F(\mathbf{r},\mathbf{p},\mathbf{q},t)=f(\mathbf{r},\mathbf{p},t)f(\mathbf{r},\mathbf{q},t)\equiv f_\mathbf{p}f_\mathbf{q}.$$
    This implies that the initial momenta of the colliding particles are independent (\textit{molecular chaos} hypothesis).
    \item In a time interval which is greater than the mean collision time and smaller than the mean free time, the distribution function varies marginally.
\end{itemize}
Having said so, if we assume the system to be homogeneous in space, i.e. if there is no dependence on the position $\mathbf{r}$, we can define:
$$f_\mathbf{p}=f(\mathbf{p},t)=\frac{1}{V}\int \dd^3 \mathbf{r}\,f(\mathbf{r},\mathbf{p},t),$$
and finally write the \textit{Boltzmann transport equation}:
$$\left(\frac{\partial f}{\partial t}\right)_{\text{coll}}=\int \dd^3 \mathbf{q}\,\dd^3 \mathbf{p}' \dd^3 \mathbf{q}'\delta^3(\mathbf{p}+\mathbf{q}-\mathbf{p}'-\mathbf{q}')\omega(\mathbf{p}\mathbf{q}\to \mathbf{p}'\mathbf{q}')[f_\mathbf{p'}f_\mathbf{q'}-f_\mathbf{p}f_\mathbf{q}],$$
where the probability amplitude $\omega(\mathbf{p}\mathbf{q}\to \mathbf{p}'\mathbf{q}')$ is defined as:
$$dP_{\mathbf{p}\mathbf{q}\to \mathbf{p}'\mathbf{q}'}\equiv \omega(\mathbf{p}\mathbf{q}\to \mathbf{p}'\mathbf{q}')d\Phi,$$
being $d\Phi$ the phase-space infinitesimal volume element available and
$dP_{\mathbf{p}\mathbf{q}\to \mathbf{p}'\mathbf{q}'}$ the probability per unit time for a transition $\mathbf{p}\mathbf{q}\to \mathbf{p}'\mathbf{q}'$ to occur.
%This can also be expressed in terms of the differential cross section:
%$$\left(\frac{\partial f}{\partial t}\right)_{\text{coll}}=\int \dd^3 \mathbf{q}\, \mathbf{v}_{\mathbf{p}\mathbf{q}}\frac{d\sigma}{d\Omega}\dd\Omega[f_\mathbf{p'}f_\mathbf{q'}-f_\mathbf{p}f_\mathbf{p}].$$
The only stationary solution at equilibrium is the so-called \textit{Jüttner-Boltzmann distribution}, it being the generalization of the Maxwell-Boltzmann function:
$$f_{\text{eq}}=\exp\left[-\frac{p^\mu u_\mu}{T}\right].$$

Moving on more specifically to the study of QGP, let us now focus on the collision term. We can introduce the quantity $\omega(\mathbf{p},\mathbf{k})$, which represents the rate of collision of a particle with momentum $\mathbf{p}$ exchanging a momentum $\mathbf{k}$: it is therefore the transition rate from a state (e.g. a heavy quark) with momentum $\mathbf{p}$ to a state with momentum $\mathbf{p}-\mathbf{k}$. In this way the collision integral, it being called $\mathcal{C}[f]$, can be written as:
\begin{equation}
    \mathcal{C}[f]=\int \dd^3 \mathbf{k}[\omega(\mathbf{p}+\mathbf{k},\mathbf{k})f(\mathbf{p}+\mathbf{k})-\omega(\mathbf{p},\mathbf{k})f(\mathbf{p})].
    \label{3.3}
\end{equation}
The expression \eqref{3.3} for the collision integral takes into account for a gain and a loss term, which in our case are due to the collisions of the heavy quark with the partons from the QGP bulk. The transition rate $\omega(\mathbf{p},\mathbf{k})$ has therefore to be considered as a sum of the scattering contributions with light quarks, light antiquarks and gluons:
$$\omega(\mathbf{p},\mathbf{k})=\omega_g(\mathbf{p},\mathbf{k})+\omega_q(\mathbf{p},\mathbf{k})+\omega_{\Bar{q}}(\mathbf{p},\mathbf{k}).$$
By focusing, for the moment, only on the contribution from gluons, the transition rate is given by \cite{Svetitsky_1988}:
\begin{equation}
    \omega_g(\mathbf{p},\mathbf{k})=\gamma_g\int \frac{\dd^3 \mathbf{q}}{(2\pi)^3}\hat{f}_g(\mathbf{q})v_{\mathbf{q},\mathbf{p}}\sigma^g_{\mathbf{p},\mathbf{q}\to \mathbf{p}-\mathbf{k},\mathbf{q}+\mathbf{k}}.
    \label{3.4}
\end{equation}
In particular, in the \eqref{3.4} it appears the distribution function of the gluons at thermodynamic equilibrium $\hat{f}_g$, the relative velocity of the colliding particles $v_{\mathbf{q},\mathbf{p}}\equiv |\mathbf{v}_\mathbf{q}-\mathbf{v}_\mathbf{p}|$, a degeneration factor $\gamma_g=2\cdot 8$ (which takes into account a sum over spin and colour degrees of freedom) and finally the differential cross section for the elastic scattering $\sigma^g$, whose expression can be put in the form:
$$\sigma^g_{\mathbf{p},\mathbf{q}\to \mathbf{p}-\mathbf{k},\mathbf{q}+\mathbf{k}}=\frac{1}{(2\pi)^6}\frac{1}{v_{\mathbf{q},\mathbf{p}}}\frac{1}{2E_\mathbf{q}}\frac{1}{2E_\mathbf{p}}\frac{1}{\gamma_g\gamma_Q}\cdot$$
$$\cdot \sum |\mathcal{M}_{gQ}|^2\frac{1}{2E_{\mathbf{p}-\mathbf{k}}}\frac{1}{2E_{\mathbf{q}+\mathbf{k}}}(2\pi)^4\delta(E_\mathbf{p}+E_\mathbf{q}-E_{\mathbf{p}-\mathbf{k}}-E_{\mathbf{q}+\mathbf{k}}).$$
The scattering matrices $\mathcal{M}$ describe the amplitude of elastic scattering. In the case of the gluons we are going to consider the contribution of the following tree-level Feynman diagrams:

\begin{fmffile}{quarkgluon}
\vspace{1em}
$$\begin{fmfgraph*}(75,50)
    \fmfbottom{i1,i2}
    \fmftop{o1,o2}
    \fmf{gluon, width=5}{o1,v1}
    \fmf{gluon, width=5}{o2,v2}
    \fmf{fermion, width=5}{i1,v1,v2,i2}
    \fmflabel{$Q$}{i1}
    \fmflabel{$Q$}{i2}
    \fmflabel{$g$}{o1}
    \fmflabel{$g$}{o2}
\end{fmfgraph*}~~~~~~~~~~~~~\begin{fmfgraph*}(75,50)
    \fmfleft{i1,i2}
    \fmfright{o1,o2}
    \fmf{fermion, tension=2, width=5}{i1,v2}
    \fmf{fermion, tension=0.6, width=5}{v2,w2}
    \fmf{fermion, tension=2, width=5}{w2,o1}
    \fmf{phantom}{w2,o2}
    \fmf{phantom}{i2,v2}
    \fmffreeze
    \fmf{gluon, width=5}{v2,o2}
    \fmf{gluon, width=5}{i2,w2}
    \fmflabel{$Q$}{i1}
    \fmflabel{$g$}{i2}
    \fmflabel{$Q$}{o1}
    \fmflabel{$g$}{o2}
\end{fmfgraph*}~~~~~~~~~~~~~\begin{fmfgraph*}(75,50)
    \fmfbottom{i1,i2}
    \fmftop{o1,o2}
    \fmf{fermion, width=5}{i1,v1,i2}
    \fmf{gluon, width=5}{v1,v2}
    \fmf{gluon, width=5}{o1,v2,o2}
    \fmflabel{$Q$}{i1}
    \fmflabel{$Q$}{i2}
    \fmflabel{$g$}{o1}
    \fmflabel{$g$}{o2}
\end{fmfgraph*}$$
\end{fmffile}
and of their interference terms. The above diagrams represent the $s$, $u$, $t$ channel processes respectively, in particular the $t$-channel contribution appears to be the dominant one. On the other hand, in the collision with the light quarks and antiquarks we have the contribution of the following $t$-channel diagram only:

\begin{fmffile}{quarkgluon}
\vspace{1.5em}
$$\begin{fmfgraph*}(83,54)
    \fmfbottom{i1,i2}
    \fmftop{o1,o2}
    \fmf{fermion, width=5}{i1,v1,i2}
    \fmf{gluon, width=5}{v1,v2}
    \fmf{fermion, width=5}{o1,v2,o2}
    \fmflabel{$Q$}{i1}
    \fmflabel{$Q$}{i2}
    \fmflabel{$q$}{o1}
    \fmflabel{$q$}{o2}
\end{fmfgraph*}$$
\vspace{.5em}
\end{fmffile}

As it is well known, the $t$ channel diagrams present a forward divergence. Since we are considering a non-zero thermal mass for the gauge bosons, we will see that such a divergence is automatically cured and we do not need the introduction of any Debye screening mass \cite{Philipsen_2001}.\\

After further manipulation, the collision integral can be written in the following form:
\begin{align}
    \mathcal{C}[f]&=\frac{1}{2E_\mathbf{p}}\int \frac{\dd^3 \mathbf{q}}{2E_\mathbf{q}(2\pi)^3}\int \frac{\dd^3 \mathbf{q'}}{2E_\mathbf{q'}(2\pi)^3}\int \frac{\dd^3 \mathbf{p'}}{2E_\mathbf{p'}(2\pi)^3}\frac{1}{\gamma_Q}\cdot\nonumber\\
    &\cdot \sum |\mathcal{M}|^2(2\pi)^4 \delta^4(p+q-p'-q')[f(\mathbf{p}')\hat{f}(\mathbf{
q}')-f(\mathbf{p})\hat{f}(\mathbf{q})],
\label{3.5}
\end{align}
being $\mathbf{p}'=\mathbf{p}-\mathbf{k}$ and $\mathbf{q}'=\mathbf{q}+\mathbf{k}$. From equation \eqref{3.5} we immediately understand that the collision integral vanishes when $f(\mathbf{p}')\hat{f}(\mathbf{
q}')=f(\mathbf{p})\hat{f}(\mathbf{q})$. This condition is satisfied when the system reaches equilibrium, therefore when both heavy quarks and partons of the bulk are described by the Jüttner-Boltzmann distribution function. Of course in this analysis we have not taken into account quantum effects like the \textit{Pauli-blocking} or the \textit{Bose-enhancement}, which take care of the fermionic and bosonic nature of the particles under consideration.

\section{The Fokker-Planck equation}
\label{The Fokker-Planck equation}
We have just described the theoretical framework of the Boltzmann equation. However, its resolution for the motion of heavy quarks is quite difficult, because of the form of $\mathcal{C}[f]$. A common approach which is used in this field to overcome this problem consists in expanding the function $\omega(\mathbf{p}+\mathbf{k},\mathbf{k})f(\mathbf{p}+\mathbf{k})$ and truncating such expansion at second order in $\mathbf{k}$, therefore assuming that $\omega(\mathbf{p},\mathbf{k})$ decreases sufficiently fast with $|\mathbf{k}|$: such assumption is acceptable when the scattering processes of the heavy quark with the bulk partons are ‘soft', i.e. there are no big momentum exchanges.\\
Under this approximation we can write:
$$\omega(\mathbf{p}+\mathbf{k},\mathbf{k})f(\mathbf{p}+\mathbf{k})\simeq \omega(\mathbf{p},\mathbf{k})f(\mathbf{p})+k_i\frac{\partial}{\partial p_i}(\omega f)+\frac12 k_i k_j \frac{\partial^2}{\partial p_i \partial p_j}(\omega f)+\dots$$
By inserting such an expansion in the expression \eqref{3.3} for the collision integral we get:
\begin{align}
    \mathcal{C}[f]&=\int \dd^3 \mathbf{k}\left[k_i \frac{\partial}{\partial p_i}(\omega f)+\frac12 k_i k_j \frac{\partial^2}{\partial p_i \partial p_j}(\omega f)\right]=\nonumber\\
    &=\frac{\partial}{\partial p_i}\left\{\int \dd^3 \mathbf{k}\, k_i \omega(\mathbf{p},\mathbf{k})f(\mathbf{p})+\frac{\partial }{\partial p_j}\left[\frac12 \int \dd^3\mathbf{k}\, k_i k_j \omega(\mathbf{p},\mathbf{k})f(\mathbf{p})\right]\right\}.\nonumber
\end{align}
If we now define:
$$A_i(\mathbf{p},T)=\int \dd^3 \mathbf{k}\,k_i\omega(\mathbf{p},\mathbf{k}),~~~~B_{ij}(\mathbf{p},T)=\frac12 \int \dd^3 \mathbf{k}\,k_ik_j\omega(\mathbf{p},\mathbf{k})$$
we can write the Boltzmann equation in the simplified form:
\begin{equation}
    \frac{d}{dt}f(\mathbf{p})=\frac{\partial}{\partial p_i}\left\{A_i(\mathbf{p},T)f(\mathbf{p})+\frac{\partial}{\partial p_j}[B_{ij}(\mathbf{p},T)f(\mathbf{p})]\right\}.
    \label{3.6}
\end{equation}
The equation \eqref{3.6} and all the equations of the above form are called \textit{Fokker-Planck equations}. The dependence on the temperature of the coefficients $A_i$ and $B_{ij}$ comes from the distribution function of the bulk partons and (in our case) also from the QPM masses which appear in the scattering matrices. By now exploiting the definition of $\omega(\mathbf{p},\mathbf{k})$ for each species (light quarks, light antiquarks, gluons) we can write the coefficient $A_i(\mathbf{p},T)$ as:
\begin{align}
&A_i(\mathbf{p},T)=\frac{1}{2E_{\mathbf{p}}}\int \frac{\dd^3 \mathbf{q}}{2E_\mathbf{q}(2\pi)^3}\int \frac{\dd^3 \mathbf{q}'}{2E_\mathbf{q'}(2\pi)^3}\int \frac{\dd^3 \mathbf{p}'}{2E_\mathbf{p'}(2\pi)^3}\frac{1}{\gamma_Q} \sum |\mathcal{M}|^2\cdot\nonumber\\
&\cdot(2\pi)^4\delta^4(p+q-p'-q')\hat{f}(\mathbf{q})[(\mathbf{p}-\mathbf{p}')_i]\equiv \langle \langle (\mathbf{p}-\mathbf{p}')_i \rangle \rangle.
\label{a}
\end{align}
Similarly for the coefficient $B_{ij}(\mathbf{p},T)$ we have:
\begin{equation}
    B_{ij}(\mathbf{p},T)=\frac12\langle \langle (\mathbf{p}-\mathbf{p}')_i(\mathbf{p}-\mathbf{p}')_j\rangle \rangle.
    \label{b}
\end{equation}
The coefficients $A_i$ and $B_{ij}$ are called \textit{drag} and \textit{diffusion coefficients}, respectively: they represent the transport coefficients for the evolution of the system. Such coefficients have a clear physical meaning, which gets manifest in the non-relativistic limit. In particular, one can see that $A_i$ is related to the drag force acting on the mean momentum of the particle, therefore it is related to its thermalization time. Instead, the coefficients $B_{ij}$ describe the diffusion processes (i.e. the fluctuations) that the particle undergoes in momentum space.\\
Since we will be mainly interested in the drag coefficient, let us notice that for an isotropic medium, rotational symmetry allows for its simplification. In particular, we can write:
$$A_i(\mathbf{p},T)=A(p,T)p_i,$$
and therefore determine the scalar quantity $A(p,T)$ as:
\begin{equation}
    A(p,T)=\frac{p_i}{\mathbf{p}^2}A_i(\mathbf{p},T)=\langle \langle 1\rangle \rangle-\langle \langle \mathbf{p}'\cdot \mathbf{p}\rangle \rangle/\mathbf{p}^2,
    \label{c}
\end{equation}
The diffusion in the momentum space can be linked to a diffusion in the coordinate space via a coefficient $D_s$, which quantifies the fluctuations in the position as:
$$\langle \mathbf{x}^2(t)\rangle-\langle \mathbf{x}(t)\rangle^2\simeq 6D_st.$$
Such a \textit{spatial diffusion coefficient} is related to the drag coefficient as follows:
\begin{equation}
    D_s=\frac{T}{M_Q\cdot A(p\to 0)}.
    \label{Ds}
\end{equation}

In order to operatively perform the calculations which allow to evaluate the drag coefficient (and therefore the $D_s$) as given in equation \eqref{c}, we have to specify a reference frame \cite{Svetitsky_1988}. In particular, for any quantity $F(\mathbf{p},\mathbf{p}',T)$ depending on the temperature and on the initial and final momenta of the heavy quark, we have previously introduced the average $\langle \langle F(\mathbf{p},\mathbf{p}',T)\rangle \rangle$ as:
\begin{align*}
\langle \langle F(\mathbf{p},\mathbf{p}',T)\rangle \rangle=&\frac{1}{2E_{\mathbf{p}}}\int \frac{\dd^3 \mathbf{q}}{2E_\mathbf{q}(2\pi)^3}\int \frac{\dd^3 \mathbf{q}'}{2E_\mathbf{q'}(2\pi)^3}\int \frac{\dd^3 \mathbf{p}'}{2E_\mathbf{p'}(2\pi)^3}\frac{1}{\gamma_Q}\cdot\\
&\cdot \sum |\mathcal{M}|^2(2\pi)^4\delta^4(p+q-p'-q')\hat{f}(\mathbf{q})F(\mathbf{p},\mathbf{p}',T).
\end{align*}

If we now work in the center of mass frame and evaluate the integral in $\dd^3 \mathbf{q}$ by placing the polar axis along the momentum $\mathbf{p}$, one can show that the previous expression reduces to \cite{Svetitsky_1988}:
\begin{align}
    \langle \langle F(\mathbf{p},\mathbf{p}',T)\rangle \rangle=\frac{1}{(2\pi)^3}\int_0^\infty \dd q\,q^2\int_{-1}^{+1}\dd\cos \alpha \int_{t_{\text{min}}}^{t_{\text{max}}}\dd t\,v_{\text{rel}}\frac{d\sigma}{dt}\cdot&\nonumber\\
    \cdot \hat{f}(q)\int_0^{2\pi}\dd\phi_{\text{CM}}F(\mathbf{p},\mathbf{p}',T)&,
    \label{4.7}
\end{align}
where $t$ is the well known Mandelstam variable, $\alpha$ is the polar angle of the integral in $\dd^3\mathbf{q}$, the differential cross section $d\sigma/dt$ is given by:
$$\frac{d\sigma}{dt}=\frac{1}{16\pi}\frac{1}{[(s-M_Q^2-m^2)^2-4M_Q^2m^2]}\frac{1}{\gamma_Q}\sum |\mathcal{M}|^2,$$
and $v_{\text{rel}}$ represents the relative speed between the heavy quark and each bulk parton (the hatted quantities are considered in the center of mass frame):
$$v_{\text{rel}}=\frac{\sqrt{s}|\hat{\mathbf{p}}|}{E_\mathbf{p}E_\mathbf{q}}.$$

As one may have noticed, the quantities appearing in the RHS of \eqref{c} represent the function $F(\mathbf{p},\mathbf{p}',T)$ to insert in \eqref{4.7}. What we are left to do is therefore to express $\mathbf{p}\cdot \mathbf{p}'$ in terms of the CM variables appearing in the integral \eqref{4.7}.\\
The Lorentz transformations from the lab frame to the CM frame are given by:
$$\hat{\mathbf{p}}=\gamma_{\text{CM}}(\mathbf{p}-\mathbf{v}_{\text{CM}}E_\mathbf{p}),~~~~\hat{E}_\mathbf{p}=\gamma_{\text{CM}}(E_\mathbf{p}-\mathbf{v}_{\text{CM}}\cdot \mathbf{p}),$$
where:
$$\gamma_{\text{CM}}=\frac{1}{\sqrt{1-|\mathbf{v}_{\text{CM}}|^2}},~~~~~\mathbf{v}_{\text{CM}}=\frac{\mathbf{p}+\mathbf{q}}{E_\mathbf{p}+E_\mathbf{q}}.$$
By fixing the axes in the CM frame, few calculations (whose details can be again found in \cite{Svetitsky_1988}) will lead to the result:
$$\mathbf{p}\cdot\mathbf{p}'=E_\mathbf{p}E_\mathbf{p}'-\hat{E}^2+\hat{\mathbf{p}}^2\cos\theta_{\text{CM}}.$$

\chapter[Shear viscosity in the CE formalism]{Shear viscosity in the Chapman-Enskog formalism}
\label{Shear viscosity in the Chapman-Enskog formalism}

\section{Viscous Hydrodynamics}
\label{Viscous hydrodynamics}
When a system reaches thermal equilibrium, at least locally, \textit{fluid dynamics} is a feasible continuum description based on the solution of macroscopic equations for the baryon number and the energy-momentum conservation laws. In a Lorentz covariant form these laws are given by:
\begin{equation}
    \partial_\mu J^\mu_B(x)=0,~~~~~\partial_\mu T^{\mu\nu}(x)=0.
    \label{2.2}
\end{equation}
In the above relations $T^{\mu\nu}$ and $J^\mu_B$ are treated as classical continuum fields, whose initial values are provided by models like Glauber \cite{Glauber_model} or CGC \cite{Lappi_2008,Fukushima_2012,MCLERRAN_2010}.\\

A \textit{perfect} or \textit{ideal fluid} is characterized by the conservation of entropy density for an observer which is locally comoving. In such Local Rest Frame (LRF), we have that the particle number current and the entropy current assume the simple forms \cite{Landau, WeinbergGC}:
$$J^\mu_B(x)=n(x)\delta_0^\mu,~~~s^\mu(x)=s(x)\delta_0^\mu,$$
being $n(x)$ and $s(x)$ the baryon number and entropy densities in the LRF, whereas the energy-momentum tensor is expressed as:
$$T^{\mu\nu}(x)=\text{diag}[\epsilon(x),P(x),P(x),P(x)].$$
Then, by applying a Lorentz transformation we get the general expressions:
\begin{align}
J^\mu_B(x)=n(x)u^\mu (x),~~~s^\mu(x)=s(x)u^\mu(x),\nonumber\\
T^{\mu\nu}(x)=[\epsilon(x)+P(x)]u^\mu (x) u^\nu (x)-P(x)\eta^{\mu\nu},
    \label{2.3}
\end{align}
being $\eta^{\mu\nu}=\text{diag}~[1,-1,-1,-1]$ the metric tensor in the Minkowski space and $u^\mu(x)$ the fluid four-velocity.\\

At this point, viscous corrections are introduced by including additional terms in the energy-momentum tensor, in the baryon current and in the entropy current, as follows:
$$T^{\mu\nu}(x)=T^{\mu\nu}_{\text{ideal}}(x)+\tau^{\mu\nu}(x),$$
$$J_B^\mu(x)=n(x)u^\mu(x)+\nu_B^\mu,$$
$$s^\mu(x)=s(x)u^\mu(x)+\sigma^\mu,$$
where $T^{\mu\nu}_{\text{ideal}}$ stands for the energy momentum tensor of the ideal fluid given in \eqref{2.3}. The particle density $n$, the energy density $\epsilon$, as well as the isotropic pressure $P$, which is related to $n$ and $\epsilon$ via the EoS, are derived exactly as they would be in the ideal case from \eqref{2.3}:
$$n(x)=u_\mu J^\mu_B,~~\epsilon(x)=u_\mu T^{\mu\nu}u_\nu, ~~P(x)=(\eta_{\mu\nu}-u_\mu u_\nu)T^{\mu\nu}/3.$$
Having done so, one can derive viscous hydrodynamic equations from the same conservation laws \eqref{2.2}. Here, however, we have an ambiguity over the definition of the fluid velocity, which is referred to as the \textit{frame problem}: if we choose the \textit{Landau-Lifshitz frame} $u^\mu$ is defined as the energy flow from $T^{\mu\nu}$ \cite{Landau}, whereas in the \textit{Eckart frame} it is defined from the particle flow $J^\mu_B$ \cite{WeinbergGC,Eckart_1940}. Regardless of the frame chosen, we obtain the following orthogonality constraints:
\begin{equation}
    u_\mu \tau^{\mu\nu}=0,~~~u_\mu \nu_B^\mu=0,
    \label{2.4}
\end{equation}
and one can get an explicit form and a physical meaning of the viscous terms by making an expansion order-by-order with respect to the derivatives of the fluid velocity.\\
For this purpose, by introducing the \textit{projector tensor} $\Delta^{\mu\nu}$, such that:
$$\Delta^{\mu\nu}=\eta^{\mu\nu}-u^\mu u^\nu,~~~ u_\mu \Delta^{\mu\nu}=0,~~~\Delta_\mu^\mu=0,$$
we can define the \textit{covariant gradient} as $\nabla^\mu\equiv \Delta^{\mu\nu}\partial_\nu$. Then, if we assume $\tau^{\mu\nu}$ and $\nu_B^\mu$ to be functions of the first derivatives $\partial_\mu u_\nu$ only, we can impose the constraints \eqref{2.4} and at first order we get:
$$\nu_B^\mu=-\frac{\mu_B}{T}\sigma^\mu=\kappa \left(\frac{nT}{\epsilon+P}\right)^2\left(\partial^\mu-u^\mu u_\nu \partial^\nu\right)\left(\frac{\mu_B}{T}\right),$$
$$\tau^{\mu\nu}=\pi^{\mu\nu}-\Pi\Delta^{\mu\nu},$$
where $\mu_B$ is the baryon chemical potential, whereas the coefficient $\kappa$ denotes the fluid heat conduction, which corresponds to the amount of energy dissipated without particle flow. Instead, the entropy equation is obtained from the identity $u_\mu \partial_\nu T^{\mu\nu}=0$, which leads to:
\begin{equation}
    \partial_\mu s^\mu=\nu_B^\mu\left(\partial_\mu-u_\mu u^\lambda \partial_\lambda\right)\left(\frac{\mu_B}{T}\right)+\frac{1}{T}\tau^{\mu\nu}\left(\partial_\mu- u_\mu u^\lambda \partial_\lambda\right) u_\nu.
    \label{2.5}
\end{equation}
Due to the second law of thermodynamics (generalized to relativistic systems) $\partial_\mu s^\mu \geq 0$, the RHS of the previous equation has to be non-negative.\\
Focusing now on the equation for $\tau^{\mu\nu}$, it can be shown that:
\begin{equation}
    \tau^{\mu\nu}=\eta \sigma^{\mu\nu}+\zeta \Delta^{\mu\nu} \nabla_\lambda u^\lambda,
    \label{2.6}
\end{equation}
in which the \textit{traceless viscous stress tensor} $\sigma^{\mu\nu}$ is being defined as:
$$\sigma^{\mu\nu}=[\partial^\mu u^\nu+\partial^\nu u^\mu-u^\lambda \partial_\lambda(u^\mu u^\nu)]-\frac23 \Delta^{\mu\nu} \nabla_\lambda u^\lambda,$$
whereas the second term in the RHS of \eqref{2.6} describes the ‘volume effect' of the viscosity, which is encoded in the trace of $\tau^{\mu\nu}$. In equation \eqref{2.6} we have the appearance of the two coefficients $\zeta,\eta \geq 0$, which are called \textit{bulk viscosity} and \textit{shear viscosity}, respectively. In particular, the latter quantity will be thoroughly studied in the following sections via the ratio $\eta/s$, being $s$ the entropy density.\\

To completely close the study of first order viscous hydrodynamics we need four other equations, coupled to both \eqref{2.6} and the EoS, which we obtain by expanding the equation $\Delta_{\mu\nu}\partial_\lambda T^{\lambda\nu}=0$. The result is the relativistic generalization of the \textit{Navier-Stokes equations} \cite{DeGroot_1980book}, which come up as the first order viscous correction of the Euler equation for dissipative fluids in the classical non-relativistic fluid dynamics. However, the introduction of linear order corrections into the entropy current violates causality, leading the solutions of $\Delta_{\mu\nu}\partial_\lambda T^{\lambda\nu}=0$ to propagate acausal information and therefore making first order Navier-Stokes viscous hydrodynamics inconsistent with the relativistic theory. This implies that in order to account for dissipative effects within a hydrodynamical description of QGP we need to go beyond the linear order. However, in this case there is no unique picture and one has to deal with the development of many theories (starting for instance from the Israel-Stewart theory \cite{Israel_1979} and ending up with the 14 moment approximation \cite{Denicol_2012}) in order to derive fluid dynamic equations which are truncated at second order and describe a causal behaviour, as desired.

\section{Chapman-Enskog expansion}

The \textit{Chapman-Enskog} (CE) \textit{expansion} is the most traditional formalism used to derive fluid dynamic quantities from the Boltzmann equation. It was originally developed for non-relativistic systems \cite{Chapman_book_1974}, but then Israel proved that it could be used with almost no modifications to describe relativistic systems as well \cite{Israel_1963}. The Chapman-Enskog formalism corresponds to the microscopic implementation of a gradient expansion, assuming that the single-particle distribution function depends only on the five primary fluid dynamical variables: temperature, chemical potential and the three independent components of the fluid velocity field (as well as their gradients). In particular, for a situation not too far from equilibrium, one may write the Lorentz invariant distribution function \cite{Wiranata_2012}:
$$f=f_0(1+\phi),$$
where the deviation function $\phi$ is assumed to be such that $|\phi|\ll 1$ and $f_0$ is the Boltzmann distribution function at local equilibrium:
$$f_0=\frac{\rho z \exp\left(u_\alpha p^\alpha/T\right)}{4\pi m^3 K_2(z)},$$
where $\rho\equiv \rho(x)$ and $T\equiv T(x)$ are the particle-number density and temperature in a proper coordinate system, $u\equiv u(x)$ is the four-velocity of the hydrodynamic particle flux (such that $u^\alpha u_\alpha=-1$) and $K_2(z)$ is the modified Bessel function of the second kind with $z\equiv m/T$. That being said, the corrections to the local distribution function (included in the term $\phi$) are then systematically arranged in terms of an expansion in powers of the Kundsen number \cite{Denicol_book_2021}.\\

As first pointed out by Grad \cite{Grad_1963}, the Chapman-Enskog expansion is an asymptotic series. In the relativistic case, sufficiently high orders in the Chapman-Enskog expansion lead to divergences of the $\phi$ term \cite{Denicol:2016bjh} and, therefore, those have very little use in the description of realistic systems \cite{Hiscock_1985}. Despite this major drawback, the Chapman-Enskog expansion is an important
development in kinetic theory and its results are useful to understand the asymptotic
behaviour of the Boltzmann equation.

\section{CE approximation for $\eta$ of a homogeneous gas}
\label{CE homogeneous gas}

By following the calculations developed in \cite{Wiranata_2012}, we can evaluate the shear viscosity $\eta$ (introduced in §\ref{Viscous hydrodynamics}) in the CE approximation for the most general case of relativistic particles of finite mass colliding with a non-isotropic and energy-dependent cross section. The results of the first and second order approximations for the shear viscosity, respectively, are given in natural units by:
\begin{align}
    \eta_I=&\frac{1}{10}T\frac{\gamma_0^2}{c_{00}},\label{4.9}\\
    \eta_{II}=&\frac{1}{10}T\frac{\gamma_0^2c_{11}-2\gamma_0\gamma_1c_{01}+\gamma_1^2c_{00}}{c_{00}c_{11}-c_{01}^2},
    \label{4.10}
\end{align}
where:
\begin{align*}
    \gamma_0=&-10\hat{h},\\
    \gamma_1=&-[\hat{h}(10z-25)-10z],\\
    c_{00}=&16\left(\omega_2^{(2)}-\frac{1}{z}\omega_1^{(2)}+\frac{1}{3z^2}\omega_0^{(2)}\right),\\
    c_{01}=&8\left[2z\left(\omega_2^{(2)}-\omega_3^{(2)}\right)+\left(-2\omega_1^{(2)}+3\omega_2^{(2)}\right)+\frac{1}{z}\left(\frac23 \omega_0^{(2)}-9\omega_1^{(2)}\right)-\frac{11}{3z^2}\omega_0^{(2)}\right],\\
c_{11}=&4\left[4z^2\left(\omega_2^{(2)}-2\omega_3^{(2)}+\omega_4^{(2)}\right)+2z\left(-2\omega_1^{(2)}+6\omega_2^{(2)}-9\omega_3^{(2)}\right)+\right.\\
    &\left.\left(\frac43 \omega_0^{(2)}-36 \omega_1^{(2)}+41 \omega_2^{(2)}\right)+\frac{1}{z}\left(-\frac{44}{3}\omega_0^{(2)}-35\omega_1^{(2)}\right)+\frac{175}{3z^2}\omega_0^{(2)}\right].
\end{align*}
In the above expressions $\hat{h}\equiv K_3(z)/K_2(z)$, being $K_n(z)$ the modified Bessel function of the second kind, whereas $\omega_i^{(s)}$ is the so-called relativistic omega integral, it being defined as \cite{Plumari_2012}:
\begin{equation}
    \omega_i^{(s)}=\frac{2\pi z^3}{K_2(z)^2}\int_1^\infty \dd y\, y^i(y^2-1)^3K_j(2zy)\int_0^\pi \dd\theta\sin\theta\,\frac{d\sigma}{d\Omega}(s,\theta)(1-\cos^s\theta).
    \label{4.11}
\end{equation}
Here $j=5/2+(-1)^i/2$, $y=\sqrt{s}/2m$ and $d\sigma/d\Omega$ is the differential cross section, which is obtained from the - properly summed and averaged - squared scattering matrix as:\footnote{Here we consider only elastic scattering.}
$$\frac{d \sigma}{d\Omega}=\frac{1}{64\pi^2 s}|\mathcal{M}|^2,$$
being $s$ is the proper Mandelstam variable.
\section{CE approximation for $\eta$ of a binary mixture}
\label{binary mixture}

The Chapman-Enskog formalism also allows to evaluate the shear viscosity of a binary mixture, whose constituents can interact with different cross sections among themselves and with one another. This formalism is going to be employed for a mixture of quarks and gluons.\\
We are going to follow the calculations from \cite{Wiranata_2013}, in which the first order CE approximation for a binary mixture is developed. Notice that also tertiary and higher component mixtures are in principle allowed in this formalism.\footnote{We have found a few mistakes in the calculations from \cite{Wiranata_2013}. Indeed, if we considered a two-component gas and put both masses and cross sections equal we would expect to find the one-component results \textit{independently on the relative concentrations}, and this did not happen in those formulas. In order to derive the right formulas we considered the $N$-component relations (again provided by \cite{Wiranata_2013}) and developed the calculations in the particular case $N=2$.}\\

To first order in the CE approximation, the shear viscosity in the case of a binary mixture is given by:
\begin{equation}
    \eta=\frac{T}{10}\frac{\gamma_1^2c_{22}+\gamma_2^2c_{11}-2\gamma_1\gamma_2c_{12}}{c_{11}c_{22}-c_{12}^2},
    \label{two_comp_eta}
\end{equation}
where $\gamma_k=-10c_k[K_3(z_k)/K_2(z_k)]$, being $K_n(z_k)$ the modified Bessel function of order $n$ and $z_k=m_k/T$ ($k=1,2$ labels the particle species).\\
The coefficients $c_{kl}$ in \eqref{two_comp_eta} are given by:
\begin{equation}
    c_{kk}=c_k^2c_{00}(z_k)+\tilde{c}_{kk}(kl),
    \label{1}
\end{equation}
\begin{equation}
    c_{kl}=\tilde{c}_{kl}(kl)~~~\text{for }k\neq l.
    \label{2}
\end{equation}
in which the concentrations $c_k$ are defined as $c_k=\rho_k/\rho$, being $\rho_k$ the mass density of each species (mass times the number density) and $\rho=\sum_k \rho_k$ the total mass density. The term $c_{00}(z_k)$ accounts for contributions from interaction between two identical particles of type $k$:
$$c_{00}(z_k)=16[\omega_2^{(2)}(z_k)-\omega_1^{(2)}(z_k)/z_k+\omega_0^{(2)}(z_k)/(3z_k^2)],$$
where $\omega_i^{(s)}$ are the \textit{relativistic omega integrals}:
$$\omega_i^{(s)}(z_k)=\frac{2\pi z_k^3}{K_2(z_k)^2}\int_0^{+\infty}d \psi\, \sinh^7\psi\, \cosh^i\psi\,K_j(2z_k\cosh\psi)\cdot$$
$$\cdot\int_0^\pi d \theta\, \sin\theta\,\sigma_{00}(\psi,\theta)(1-\cos^s\theta).$$
Here $\sigma_{00}$ is the differential cross section for interaction between two particles of the same species, and:
$$\cosh\psi=\frac{\sqrt{s}}{2m}=\frac{\sqrt{(p_1+p_2)^2}}{2m_k},~~~j=\frac{5}{2}+\frac{1}{2}(-1)^i.$$
Instead, in \eqref{1} and \eqref{2} the coefficients $\tilde{c}_{kl}$ and $\tilde{c}_{kk}$ are used to denote contributions to the shear viscosity due to interaction between different particle species:
$$\tilde{c}_{12}=\frac{32\rho^2c_1^2c_2^2}{3M_{12}^2n^2x_1x_2}[-10z_1z_2\zeta_{12}^{-1}Z_{12}^{-1}\omega_{1211}^{(1)}(\sigma_{12})-10z_1z_2\zeta_{12}^{-1}Z_{12}^{-2}\omega_{1311}^{(1)}(\sigma_{12})+$$
$$3\omega_{2100}^{(2)}(\sigma_{12})-3Z_{12}^{-1}\omega_{2200}^{(2)}(\sigma_{12})+Z_{12}^{-2}\omega_{2300}^{(2)}(\sigma_{12})],$$

$$\tilde{c}_{11}=\frac{32\rho^2c_1^2c_2^2}{3M_{12}^2n^2x_1x_2}[10z_1^2\zeta_{12}^{-1}Z_{12}^{-1}\omega_{1220}^{(1)}(\sigma_{12})+10z_1^2\zeta_{12}^{-1}Z_{12}^{-2}\omega_{1320}^{(1)}(\sigma_{12})+$$
$$3\omega_{2100}^{(2)}(\sigma_{12})-3Z_{12}^{-1}\omega_{2200}^{(2)}(\sigma_{12})+Z_{12}^{-2}\omega_{2300}^{(2)}(\sigma_{12})],$$

$$\tilde{c}_{22}=\frac{32\rho^2c_1^2c_2^2}{3M_{12}^2n^2x_1x_2}[10z_2^2\zeta_{12}^{-1}Z_{12}^{-1}\omega_{1202}^{(1)}(\sigma_{12})+10z_2^2\zeta_{12}^{-1}Z_{12}^{-2}\omega_{1302}^{(1)}(\sigma_{12})+$$
$$3\omega_{2100}^{(2)}(\sigma_{12})-3Z_{12}^{-1}\omega_{2200}^{(2)}(\sigma_{12})+Z_{12}^{-2}\omega_{2300}^{(2)}(\sigma_{12})].$$

In the above relations $x_k=n_k/n$, where $n_k$ is the particle number density of particle type $k$, and $n=\sum_k n_k$ is the total particle number density. We have labeled by $M_{12}=m_1+m_2$ the summed mass and by $\mu_{12}=m_1 m_2/M_{12}$ the reduced mass. Moreover, $\zeta_{12}=2\mu_{12}/T$ and $Z_{12}=M_{12}/(2T)$.\\
The expression for the relativistic omega integrals $\omega_{rtuv}^{(s)}$ is:
$$\omega_{rtuv}^{(s)}=\frac{\pi \mu_{12}}{4T K_2(z_1)K_2(z_2)}\int_0^{+\infty}d \Psi_{12}\sinh^3\Psi_{12}\cdot \left(\frac{g_{12}^2}{2\mu_{12}T}\right)^r\left(\frac{M_{12}}{P_{12}}\right)^t\cdot$$
$$\cdot (\cosh \psi_1)^u (\cosh \psi_2)^v\cdot K_\nu \left(\frac{P_{12}}{T}\right)\int_0^\pi d\theta \sin \theta \,\sigma_{12}(\Psi_{12},\theta)(1-\cos^s\theta),$$
where $\sigma_{12}$ is the differential cross section for interaction among particles of the two species, and:
$$P_{12}^2=m_1^2+m_2^2+2m_1m_2\cosh\Psi_{12},~~~g_{12}=\frac{m_1m_2\sinh\Psi_{12}}{P_{12}},$$
$$\Psi_{12}\equiv \psi_1+\psi_2,~~~\nu=\frac{5}{2}-\frac{1}{2}(-1)^{t+u+v},$$
$$\cosh\psi_1=\frac{1}{P_{12}}(m_1+m_2\cosh\Psi_{12}),~~~\cosh\psi_2=\frac{1}{P_{12}}(m_2+m_1\cosh\Psi_{12}).$$
Of course, $P_{12}=\sqrt{(p_1+p_2)^2}$ is the invariant center-of-mass energy of the two particles colliding with initial four momenta $p_1$ and $p_2$.\\

In our work the relative mass concentrations $c_k$ and the particle number concentrations $x_k$ have been obtained via a Boltzmann distribution:
$$x_1(T)=\frac{d_g \int \frac{\dd^3 \mathbf{p}}{(2\pi)^3}\exp{-\frac{\sqrt{p^2+m_g^2}}{T}}}{\sum_{i=g,q} d_i\int \frac{\dd^3 \mathbf{p}}{(2\pi)^3}\exp{-\frac{\sqrt{p^2+m_i^2}}{T}}},$$

$$c_1(T)=\frac{m_g\cdot d_g \int \frac{\dd^3 \mathbf{p}}{(2\pi)^3}\exp{-\frac{\sqrt{p^2+m_g^2}}{T}}}{\sum_{i=g,q} m_i \cdot  d_i\int \frac{\dd^3 \mathbf{p}}{(2\pi)^3}\exp{-\frac{\sqrt{p^2+m_i^2}}{T}}},$$

$$x_2(T)=1-x_1(T), ~~~ c_2(T)=1-c_1(T).$$
where $d_g=2\cdot(N_c^2-1)=16$ is the degeneration factor of gluons (spin $\times$ colour) and $d_q=2\cdot N_c \cdot  2 \cdot  N_f=36$ is the degeneration factor of quarks (spin $\times$ colour $\times$ antiparticles $\times$ flavours).\\

Of course, when the masses of the two components and all the cross sections are equal we obtain the one component results back, regardless the concentrations $c_k$. 

\chapter[Shear viscosity in the GK formalism]{Shear viscosity in the Green-Kubo formalism}
\label{Shear viscosity from the Green-Kubo formalism}

As for other transport coefficients, like heat-conductivity and bulk viscosity, also the shear viscosity $\eta$ can be related to the correlation function of the corresponding flux or tensor at thermal equilibrium \cite{Green_1954,Kubo_1957}. In particular, it is possible to derive these dissipative coefficients from the microscopic theory, by using
linear response theory: the underlying physical reason is that dissipation of fluctuations has the same physical origin as the relaxation towards equilibrium, therefore both dissipation and relaxation time are determined by the same transport coefficients \cite{Plumari_2012}.\\
In this context the expression we obtain for the shear viscosity is the \textit{Green-Kubo formula} \cite{zubarev1996statistical}:
\begin{equation}
    \eta=\frac{1}{T}\int_0^{+\infty}\dd t \int_V \dd^3 \mathbf{x} \langle \pi^{xy}(\mathbf{x},t)\pi^{xy}(\mathbf{0},t)\rangle,
    \label{4.1}
\end{equation}
where $T$ is the temperature, $\pi^{xy}$ is the $xy$ matrix element of the shear component of the energy momentum tensor, while $\langle ... \rangle$ denotes the ensemble average.\\

Here we want to determine the above integrand $\langle \pi^{xy}(\mathbf{x},t)\pi^{xy}(\mathbf{0},t)\rangle$ by employing transport simulations for a particle system in a static box of volume $V$ at equilibrium, using a proper relativistic transport code. We consider a system consisting of classical, ultra-relativistic particles which are interacting via two-body collisions. The starting point is the relativistic Boltzmann-Vlasov equation:
\begin{equation}
    p^\mu \partial_\mu f(x,p)+M(x)\partial_\mu M(x) \partial^\mu_p f(x,p)=\mathcal{C}(x,p),
    \label{4.2}
\end{equation}
where $f(x,p)$ is the distribution function for on-shell particles and $\mathcal{C}(x,p)$ is the Boltzmann-like collision integral:
\begin{equation}
\mathcal{C}(x,p)=\int_2 \int_{1'} \int_{2'} (f_{1}f_{2}-f_{1'}f_{2'})|\mathcal{M}_{1 2\to 1'2'}|^2(2\pi)^4\delta^4(p_1+p_2-p_1'-p_2')
    \label{4.3}
\end{equation}
in which $\int_j=\int \dd^3 p_j/[(2\pi)^32E_j]$, $f_j$ are the particle distribution functions and $\mathcal{M}$ denotes the transition matrix for the elastic processes, which are directly linked to the differential cross section via:
$$\frac{d\sigma}{d\Omega}=\frac{1}{64\pi^2 s}|\mathcal{M}|^2,$$
being $s$ the Mandelstam invariant. The following calculation of the shear viscosity will be carried out via a stochastic implementation of the collision integral \eqref{4.3}.

\section{The test-particles method}
In order to solve the transport equation we have used the \textit{test particle method} \cite{Wong_1982}, which basically consists in sampling the phase space distribution function using a large number of test particles. These are usually chosen as point-like, i.e. as delta functions in both coordinate and momentum space. With these assumptions the phase space distribution can be written as:
$$f(\mathbf{x},\mathbf{p},t)=A\cdot \sum_{i=1}^{N_{\text{test}}}\delta^3(\mathbf{x}-\mathbf{x}_i(t))\delta^3(\mathbf{p}-\mathbf{p}_i(t)).$$
In the above, $\mathbf{x}_i(t)$ and $\mathbf{p}_i(t)$ are respectively the position and the momentum of the $i$-th test particle, $N_{\text{test}}$ is the total number of test particles while $A$ is a normalization factor that is related to the total number of particles in a way that the integral over the phase space of the distribution function is equal to the total number of ‘physical' particles $N_{\text{particles}}$:
$$\int \dd^3 \mathbf{x} \int \frac{\dd^3\mathbf{p}}{(2\pi)^3}f(\mathbf{x},\mathbf{p},t)=\frac{A}{(2\pi)^3}N_{\text{test}}=N_{\text{particles}}.$$
This means that $(2\pi)^3/A$ turns out equal to the number of test particles over the number of real particles.\\
For a multi-species mixture, as we will deal with in our work, the relative abundance of each species will be distributed according to a Boltzmann distribution $\propto \exp{-m/T}$.\\

At this point, the study of the transport equation reduces to the solution the classical equations of motion for the test particles, each satisfying the mass-shell relation $p_\mu p^\mu=m^2$. Indeed, it can be shown, using the Liouville theorem \cite{Liouville_1838}, that the phase space distribution given as a collection of point-like test particles is a solution of the Boltzmann-Vlasov equation \eqref{4.2} if the positions and the momenta of test particles obey the relativistic Hamilton equations:
$$\dot{\mathbf{x}}_i=\frac{\mathbf{p}_i}{E_i},~~~~\dot{\mathbf{p}}_i=-\nabla_\mathbf{x}E_i+\text{coll}.$$
The term 'coll' indicates the effect of the collision integral, whose numerical implementation will be described in the following subsection. Numerically, the equations of motion have been solved using the $4^{th}$ order Runge-Kutta method.
\section{The stochastic method}
Let us now discuss the numerical implementation of the collision integral \eqref{4.3}, as first developed in \cite{Xu_2005, Lang_1993}. In this method we associate a probability collision $P$ to each pair of particles: only particles being in the same cell can collide with each other and if such probability $P$ is greater than a random number between 0 and 1 the collision takes place.\\

An expression for $P$ can be derived from the collision term of the Boltzmann-Vlasov equation \eqref{4.3}. As a stating point, the probability per unit volume $\Delta^3 \mathbf{x}$ and unit time $\Delta t$ can be defined as the ratio between the number of collisions that happen in such volume $\Delta^3 \mathbf{x}$ during the time $\Delta t$ and the total number of pairs present in that unit volume (we will only consider $2\to 2$ processes):
$$P=\frac{\Delta N_{\text{coll}}^{2\to2}}{\Delta N_1 \Delta N_2}.$$
Here $\Delta N_{\text{coll}}^{2\to2}$ is derived from the discretized form of the collision integral in \eqref{4.3}. In particular, let us consider any two particles in a spatial volume element $\Delta^3 \mathbf{x}$ with momenta in the range $(\mathbf{p}_1,\mathbf{p}_1+\Delta^3 \mathbf{p}_1)$ and $(\mathbf{p}_2,\mathbf{p}_2+\Delta^3 \mathbf{p}_2)$. The collision rate per unit phase space for such particle pair can be derived from \eqref{4.3} as \cite{Xu_2005}:
\begin{align*}
    \frac{\Delta N_{\text{coll}}^{2\to2}}{\Delta t \frac{1}{(2\pi)^3}\Delta^3 \mathbf{x}\Delta^3 \mathbf{p}_1}=\frac{1}{2E_1}&\frac{\Delta^3 \mathbf{p}_2}{(2\pi)^32E_2}f_1f_2\cdot\\
    &\int_{1'} \int_{2'}|\mathcal{M}_{1 2\to 1'2'}|^2(2\pi)^4\delta^4(p_1+p_2-p_1'-p_2').
\end{align*}
Expressing the distribution functions as:
$$f_i=\frac{\Delta N_i}{\frac{1}{(2\pi)^3}\Delta^3 \mathbf{x} \Delta^3 \mathbf{p}_i}~~~i=1,2;$$
and employing the usual definition of cross section \cite{DeGroot_1980book}:
\begin{align*}
    \sigma_{\text{tot}}=&\frac{1}{2\sqrt{[s-(m_1+m_2)^2][s-(m_1-m_2)^2]}}\cdot\\
    &\cdot\int_{1'}\int_{2'}|\mathcal{M}_{1 2\to 1'2'}|^2(2\pi)^4\delta^4(p_1+p_2-p_1'-p_2'),
\end{align*}
one obtains the absolute collision probability in a unit box $\Delta^3 \mathbf{x}$ and unit time $\Delta t$:
$$P=v_{\text{rel}}\sigma_{\text{tot}}\frac{\Delta t}{\Delta^3 \mathbf{x}},$$
where the relative velocity is defined as:
$$v_{\text{rel}}=\frac{\sqrt{[s-(m_1+m_2)^2][s-(m_1-m_2)^2]}}{2E_1E_2}$$
and $s=(p_1+p_2)^2$ is the well known Mandelstam variable. Since we use the test particle method the cross section has to be rescaled by $N_{\text{test}}$, therefore we get:
$$P=v_{\text{rel}}\frac{\sigma_{\text{tot}}}{N_{\text{test}}}\frac{\Delta t}{\Delta^3 \mathbf{x}}.$$

In the limit $\Delta t\to 0$ and $\Delta^3 \mathbf{x}\to 0$, $P$ is Lorentz invariant and the stochastic method converges to the exact solution of the Boltzmann equation. Of course the space-time discretization has to be chosen smaller than the typical scales of spatial and temporal inhomogeneities. Moreover, in practice, one should choose suitable $\Delta^3 \mathbf{x}$ and $\Delta t$ in order to make $P$ consistently less than 1.\\
The evolution of particles is therefore intuitively straightforward: particles move along straight lines between two collision events. Then, after a particular collision the momenta of colliding particles are changed statistically according to the differential cross section \cite{Xu_2005}. Strictly speaking, such 'collisions' have not to be considered as real ones, but rather as a way to map the evolution in the phase space as induced by the matrix element $\mathcal{M}$, which is stochastically sampled.\\

\section{Discretization procedure}

The shear component of the energy-momentum tensor is given by \cite{Plumari_2012}:
$$\pi^{xy}(\mathbf{x},t)=T^{xy}(\mathbf{x},t)=\int \frac{\dd^3 \mathbf{p}}{(2\pi)^3}\frac{p^xp^y}{E}f(\mathbf{x},\mathbf{p},t),$$
where we notice that the $xy$ component of the shear stress tensor, at equilibrium, is given by the energy-momentum tensor itself. Since we need the volume-averaged shear tensor, for an homogeneous system this can be written as:
$$\pi^{xy}(t)=\frac{1}{V}\sum_{i=1}^{N_{\text{test}}}\frac{p_i^xp_i^y}{E_i},$$
where $i$ runs over all the particles in the box.\\

In other works the shear viscosity is obtained via an exponential fit of the correlator $\langle \pi^{xy}(t) \pi^{xy}(0)\rangle$, i.e. by identifying the proper $\tau$, characteristic of the exponential decay, and then linking such constant to $\eta$ \cite{Plumari_2012}. Instead, we will derive the shear viscosity by discretizing the integral in \eqref{4.1} as follows:
$$\eta=\frac{V}{T}\left\langle \Delta t\sum_{j=1}^{N_{T}}\pi^{xy}(j\Delta t) \pi^{xy}(0)\right\rangle,$$
where $N_{T}=T_{\text{max}}/\Delta t$, $T_{\text{max}}$ is the maximum time chosen in our simulation and $\langle ...\rangle$ denotes the ensemble average, which is here obtained as an average over events generated numerically. Of course $T_{\text{max}}$ has to replace infinity in \eqref{4.1}, therefore it must be chosen as a sufficiently high number in order to ensure convergence.\\

\begin{figure}
    \centering
    \includegraphics[scale=0.5]{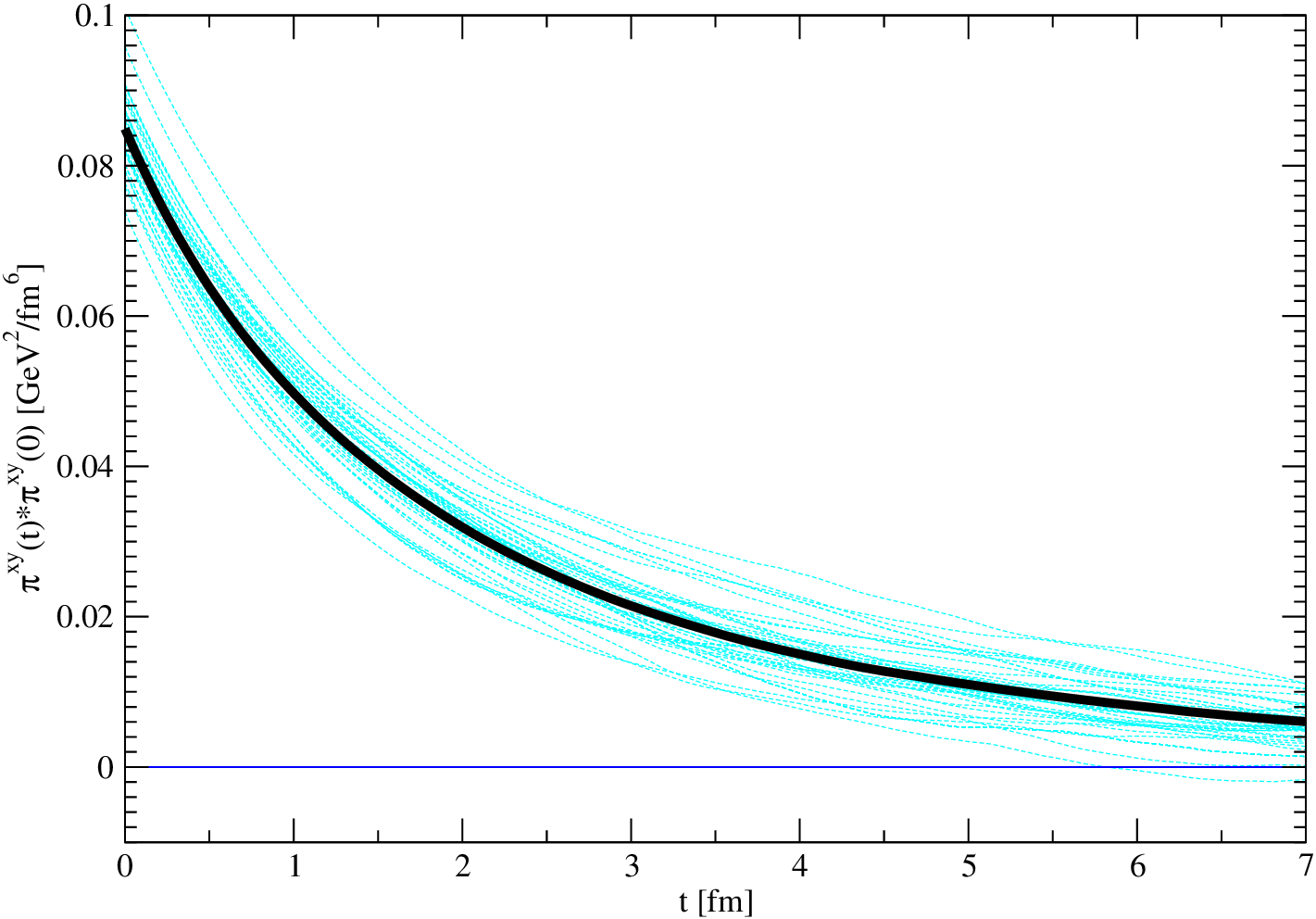}
    \caption{The behaviour of $\pi^{xy}(t) \pi^{xy}(0)$ with respect to $t$ for a temperature $T=0.5$ GeV in the physical case (see §\ref{Perturbative scattering of massive partons} and Appendix \ref{Appendix A}). In cyan we show the results for 45 different numerical events, in black their average, which highlights a clear exponential behaviour.}
    \label{fig4.1}
\end{figure}

As an example, in Figure \ref{fig4.1} we show 45 different calculations for the time behaviour of $\pi^{xy}(t) \pi^{xy}(0)$ at a temperature of $T=0.5$ GeV (cyan curves): those have been performed in the ‘physical case' in which we considered massive quarks and massive gluons, colliding with tree level pQCD scattering matrices (as already mentioned in §\ref{Perturbative scattering of massive partons} and then further developed in Appendix \ref{Appendix A}). Together with the results of each single event, which carry a significant amount of fluctuations, we show their ensemble average (in black), highlighting the typical exponential decay.

\chapter{Results}

In this chapter we show the results obtained for various transport coefficients as obtained in the QPM. More specifically, after performing a comparison between the results from Green-Kubo and the results from Chapman-Enskog for various simple cases, we will evaluate $\eta/s$ and $D_s$. We will then comment on our results and make comparisons with the existing literature. Moreover, we will discuss the temperature dependence of the ratio $(2\pi TD_s)/(4\pi \eta/s)$ and its importance in order to distinguish strong and weak coupling regimes in Hot QCD.\\

The evaluation of the Green-Kubo correlator in the relativistic Boltzmann equation has been performed employing a simulation code which has been developed in the theoretical group of INFN-LNS \& DFA-UniCT.

\section{Setting up the Green-Kubo method}
\label{Setting up the Green-Kubo method}

Before dealing with the physical case (that is, a mixture of quarks and gluons interacting via proper pQCD matrices), we set up the transport code to check the accordance with CE in some simpler cases:
\begin{itemize}
    \item[a)] Light quarks (u, d, s, of equal QPM mass $m_q$ as in \eqref{3.9}) and gluons colliding with a constant differential cross section, uniform for all processes;
    \item[b)] Light quarks and gluons colliding with 5 constant differential cross sections, for each process $qq'\to qq'$, $q\Bar{q}\to q\Bar{q}$, $qq\to qq$, $qg\to qg$, $gg\to gg$;
    \item[c)] Only gluons colliding with a pQCD-like simplified differential cross section;
    \item[d)] Only gluons colliding with the full pQCD matrix element as discussed in §\ref{Perturbative scattering of massive partons} and Appendix \ref{Appendix A}.
\end{itemize}

In order to do so, the transport code has been developed on a cubic box of volume $V=(5.2)^3$ fm with periodic boundary conditions, discretized in a lattice whose spacing is $\Delta x=\Delta y=\Delta z=0.4$ fm: those values ensure convergence within a reasonable running time. Each event would run with a total number of test particles (that is the number of real particles times the number of test particles for each real particle) of about 1-1.5 million. The total time has been empirically fixed large enough that the exponential behaviour of each correlator is appropriately dumped, but not too large in order to exclude the noise which arises from the statistical fluctuations of each event. Indeed, those fluctuations occur massively when the correlator approaches its vanishing value at late times. Also the time step has been empirically determined, by assessing the convergence of the correlator itself.\\

\begin{figure}
      \begin{subfigure}{0.49\textwidth}
     \includegraphics[scale=0.38]{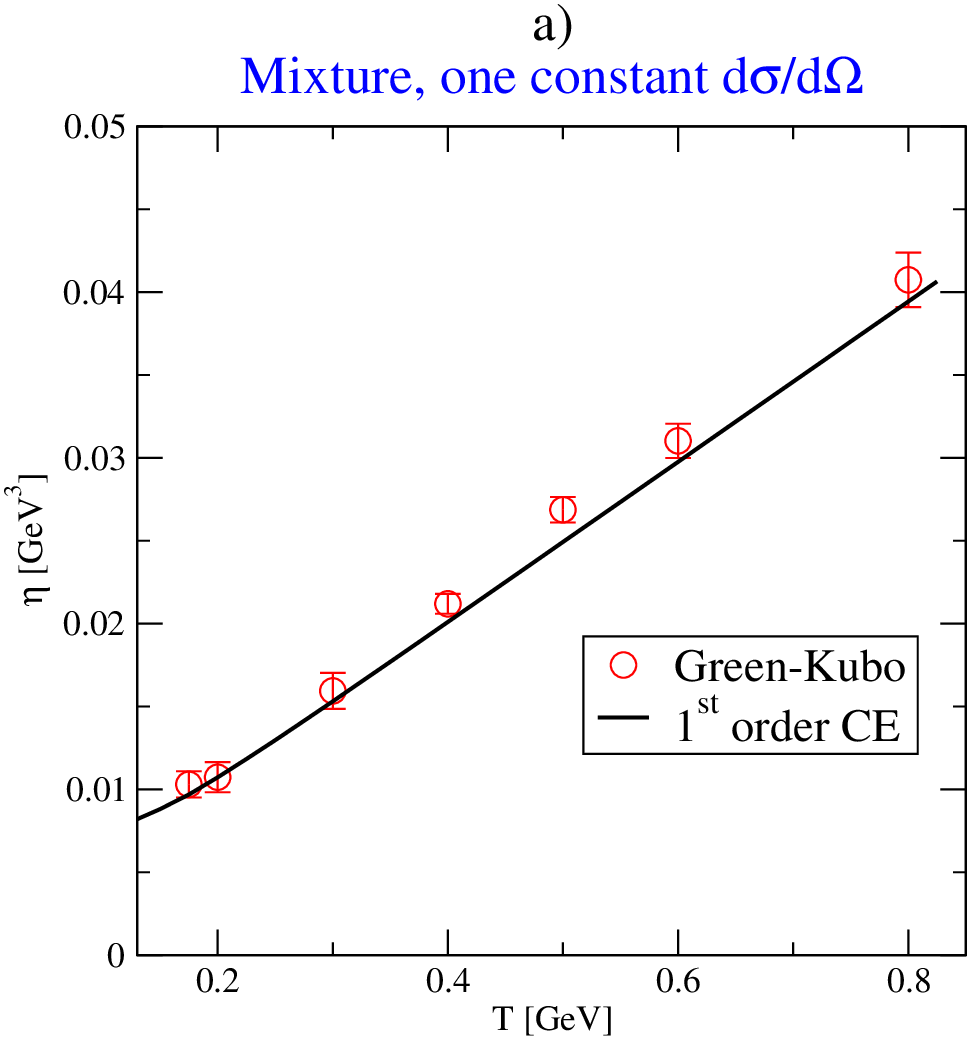}
    %\caption{Caption}
 \end{subfigure}
 \hfill
 \begin{subfigure}{0.49\textwidth}
     \includegraphics[scale=0.38]{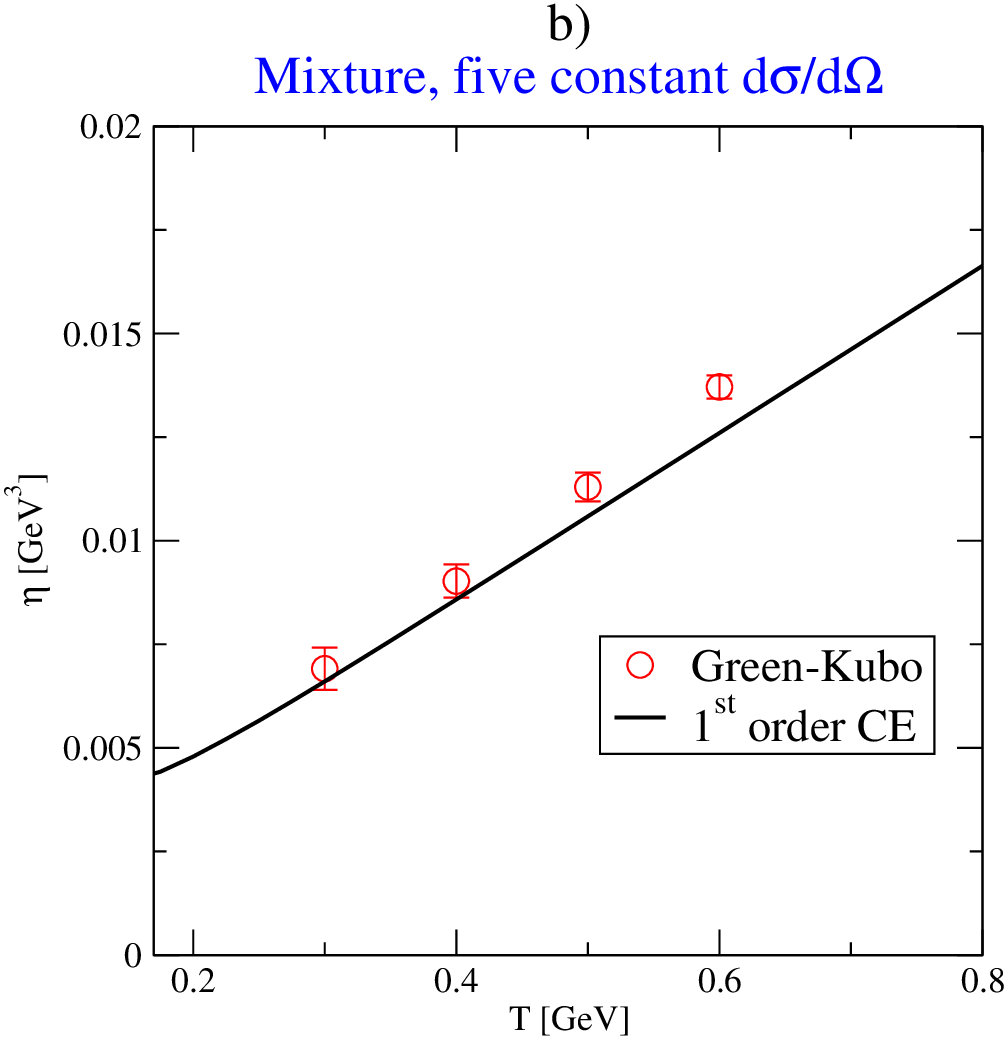}
    %\caption{Caption}
 \end{subfigure}

\medskip

\begin{subfigure}{0.49\textwidth}
     \includegraphics[scale=0.38]{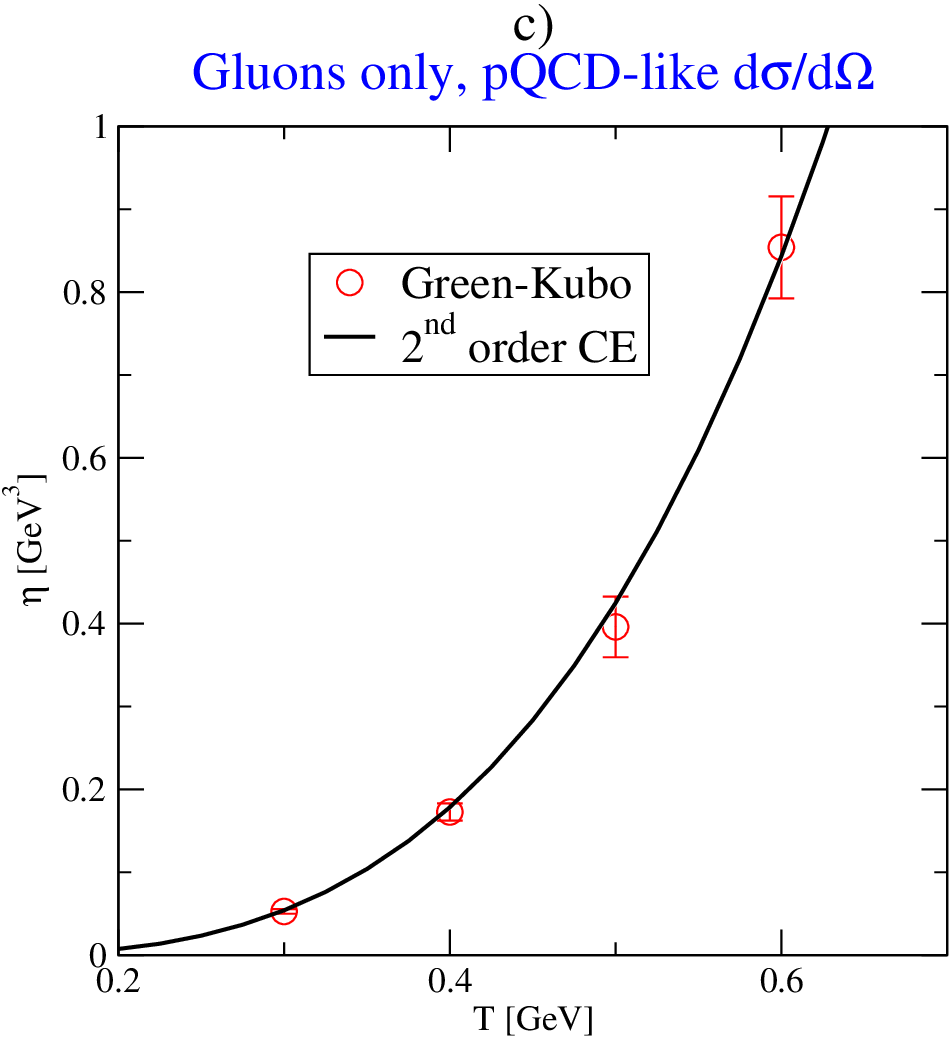}
    %\caption{Caption}
 \end{subfigure}
 \hfill
 \begin{subfigure}{0.49\textwidth}
     \includegraphics[scale=0.38]{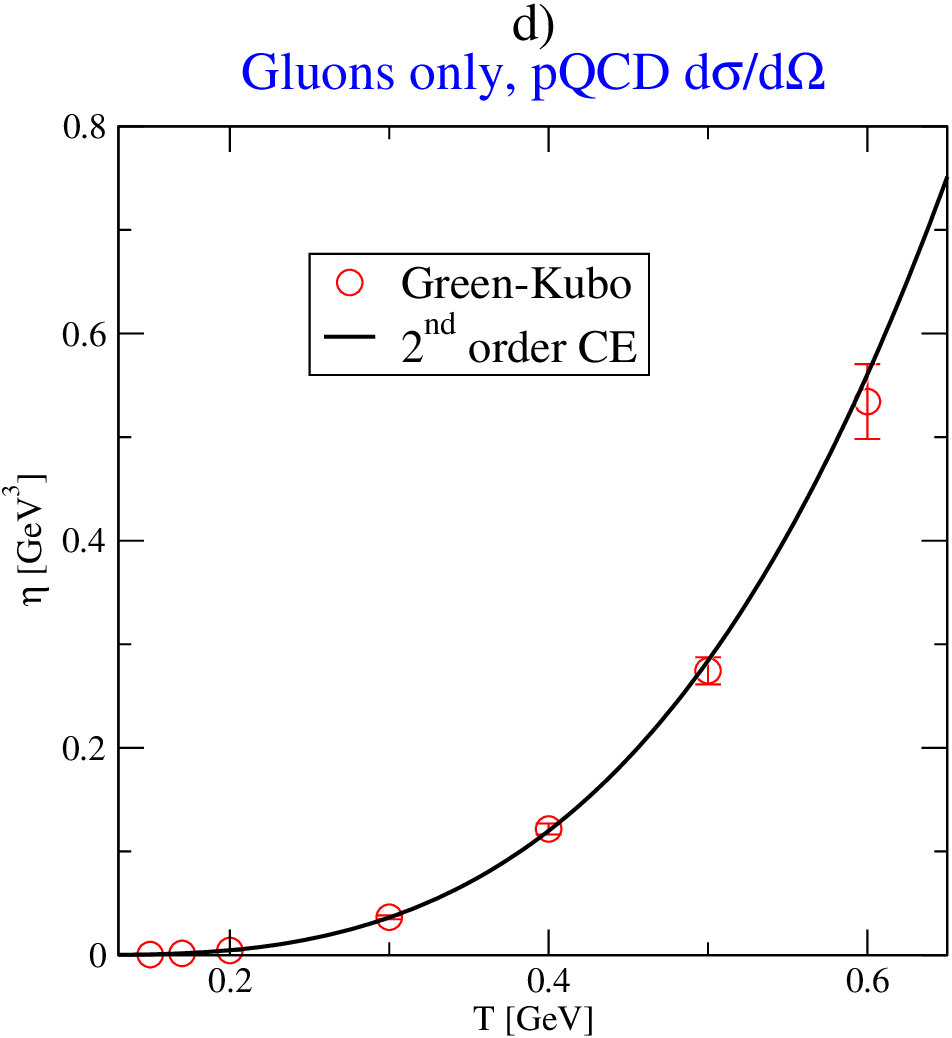}
    %\caption{Caption}
 \end{subfigure}
 \caption{We compared the results from Green-Kubo with the ones from Chapman-Enskog for four different `test cases'. For further details see the main text.}
\label{fig5.1}
 \end{figure}

We show the results in Figure \ref{fig5.1}, each set of Green-Kubo data has been obtained by averaging over 45 events and then compared with the analogous results from the Chapman-Enskog method. For the particular cases of the gluon-only events (i.e. the homogeneous case) the results obtained in the $2^\text{nd}$ order Chapman-Enskog approximation are available: in these cases we do not also show the results from the $1^\text{st}$ order, since those are nearly indistinguishable from the ones from the $2^\text{nd}$ order.\\
%(*partition step of Tmax: for T=0.6 we have 600=6000/10, for T=0.175 \
%we have 20000/10=2000, otherwise 60000/10=6000*)

Referring to Figure \ref{fig5.1}, the tests have been performed as follows:

\begin{itemize}
    \item[a)]  A mixture of light quarks and gluons colliding with an isotropic (constant angle-independent) differential cross section, fixed for all processes equal to $d\sigma/d\Omega=1/4\pi$ fm$^2$.
    \item[b)] A mixture of light quarks and gluons colliding with an isotropic (constant angle-independent) differential cross section, whose values have been chosen as:\footnote{Notice that the 3 quark-quark cross sections have been chosen as quite different from one another. This has been done in order to check the validity of the ‘average' procedure which we are illustrating quite soon.}
    $$\frac{d\sigma}{d\Omega}(qq'\to qq')=\frac{0.1}{4\pi}\text{ fm$^2$}, ~~\frac{d\sigma}{d\Omega}(q\Bar{q}\to q\Bar{q})=\frac{1}{4\pi}\text{ fm$^2$},$$
    $$\frac{d\sigma}{d\Omega}(qq\to qq)=\frac{10}{4\pi}\text{ fm$^2$}, ~~\frac{d\sigma}{d\Omega}(qg\to qg)=\frac{3}{4\pi}\text{ fm$^2$},$$
    $$\frac{d\sigma}{d\Omega}(gg\to gg)=\frac{3}{4\pi}\text{ fm$^2$}.$$
    The transport code which has been employed to evaluate the Green-Kubo correlator allows for the different quark cross sections to be implemented singularly for each different channel. On the contrary, our Chapman-Enskog formalism for the binary mixture is not able to distinguish among the different quark processes. Due to this fact, in the CE formalism we considered an 'averaged' quark-quark differential cross section as follows:
    \begin{equation}
        \frac{d\sigma}{d\Omega}=\frac16 \frac{d\sigma}{d\Omega}(qq\to qq)+\frac16\frac{d\sigma}{d\Omega}(q\Bar{q}\to q\Bar{q})+\frac{2}{3}\frac{d\sigma}{d\Omega}(qq'\to qq').
        \label{5.1}
    \end{equation}
    The factors appearing in the \eqref{5.1} refer to the 'probability' that in a collision among two quarks we have a pair of equal quarks, a quark-antiquark pair and all the other possible cases, respectively. Those have been derived via basic combinatorics, supposing a very large number of particles $N$, as:\footnote{Since the relative abundance in the mixture only depends on the mass, all quarks and antiquarks appear in equal number. This implies that the above ‘probabilities' are fixed and do not depend on either temperature nor quark mass.}
    $$\mathrm{P}_{qq\to qq}=\frac{\text{\# of possible $qq$ pairs}}{\text{total \# of possible pairs}}=\frac{N(N/6-1)/2}{N(N-1)/2}\to\frac16,$$
    $$\mathrm{P}_{q\Bar{q}\to q\Bar{q}}=\frac{\text{\# of possible $q\Bar{q}$ pairs}}{\text{total \# of possible pairs}}=\frac{N(N/6)/2}{N(N-1)/2}\to\frac16,$$
    $$\mathrm{P}_{qq'\to qq'}=1-\mathrm{P}_{qq\to qq}-\mathrm{P}_{q\Bar{q}\to q\Bar{q}}=\frac{2}{3}.$$
    \item[c)] A bulk of only gluons colliding with the following (angle-dependent) `pQCD like' differential cross section \cite{Plumari_2012}:
    $$\frac{d\sigma^{gg\to gg}}{dq^2}=9\pi \alpha_s^2 \frac{1}{(t-m_D^2(T))^2},$$
    where $m_D(T)=g(T)T$ is the Debye screening mass \cite{Philipsen_2001}.
    \item[d)] A bulk of only gluons colliding with the pQCD differential cross section, as derived in \ref{Perturbative scattering of massive partons} and Appendix \ref{Appendix A}.
\end{itemize}

The results in Figure \ref{5.1} show the agreement of the CE results (black line) with the GK results (red dots): for the latter the uncertainty has been evaluated from the statistical error arising from the average over 45 events, i.e. as the standard deviation over the mean value. We also see that, for the cases a) and b), at high temperatures the Green-Kubo averages for $\eta$ are slightly higher than the corresponding CE values with a deviation of less than $ 8 \%$. This may be due to the poor convergence of the CE expansion, in particular for a binary mixture we see that the $1^\text{st}$ order CE expansion does not provide a sufficiently accurate approximation for $T\gtrsim$ 500 MeV. Instead, we see that this does not happen in the cases c) and d) (we already mentioned that the $1^{\text{st}}$ and the $2^{\text{nd}}$ order CE data are seen to be almost equal in those cases), hence we can guess that the CE expansion converges more slowly for binary mixtures than it does in the homogeneous case.

\section{Shear viscosity $\eta/s$}
Let us now show the results for the ratio $\eta/s$, where $s$ is the entropy density which is simply evaluated in the QPM as:
$$s=\frac{\epsilon_{qp}+P_{qp}}{T},$$
$\epsilon_{qp}$ and $P_{qp}$ being given by equations \eqref{3.11} and \eqref{P_QPM}, respectively.\\

\begin{figure}[H]
    \centering
\includegraphics[width=0.85\textwidth]{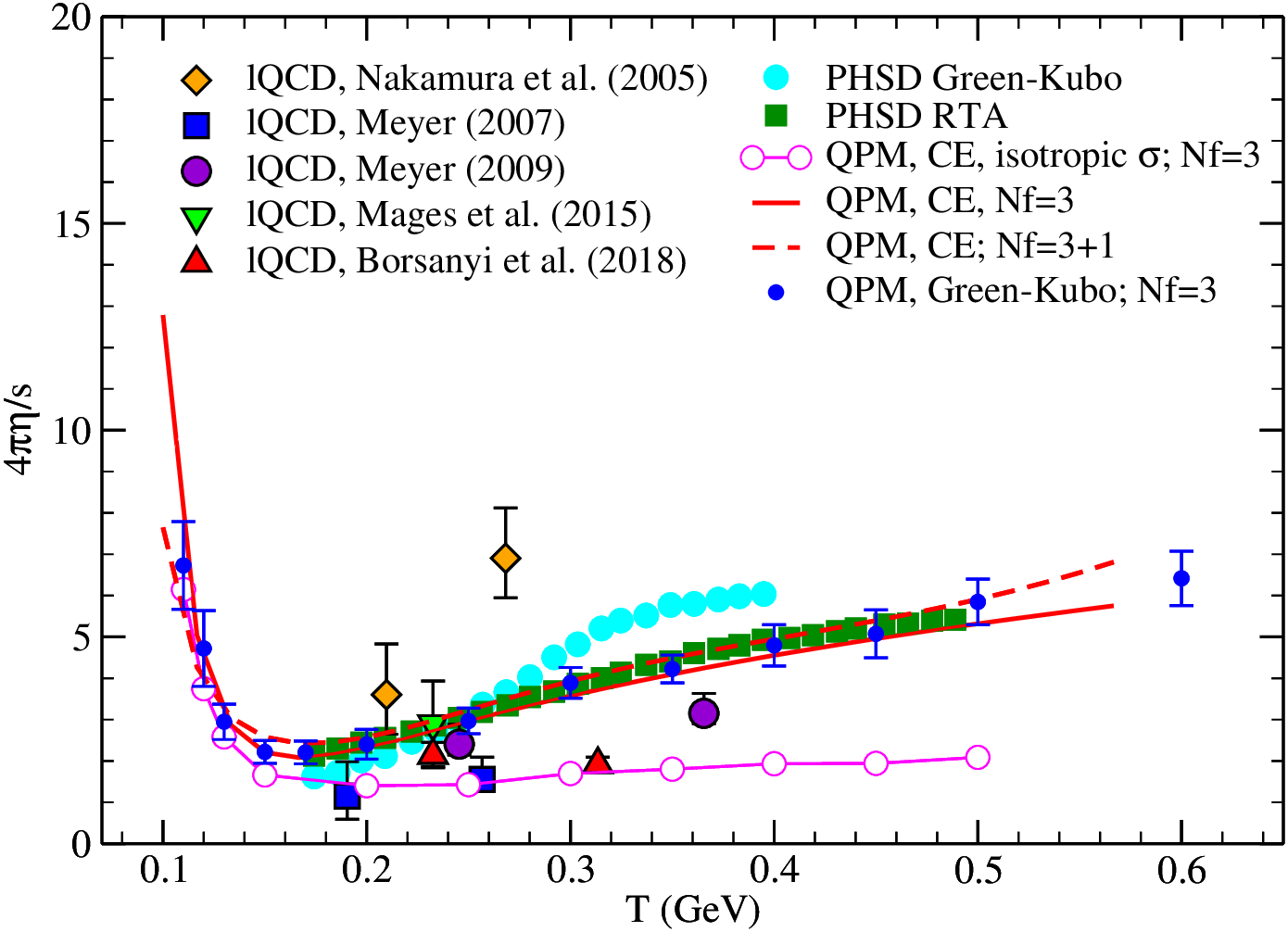}
    \caption{The ratio shear viscosity over entropy density $4\pi \eta/s$ in the CE approximation (full and dashed red lines, depending on the number of flavours), compared to the GK averages (blue circles) and the CE data in the ‘isotropic' reduction (pink empty circles, see text). Together with our results we also show the data from the PHSD group, with two different ways of calculating the $\eta$ \cite{Ozvenchuk_2013}. The other various symbols stand for different lQCD calculations, from \cite{Nakamura_2005} (orange losanges), \cite{Meyer_2007} (blue squares), \cite{Meyer_2009} (purple circles), \cite{Mages_2015} (light green downward triangle) and \cite{Borsanyi_2018} (red upward triangles).}
    \label{eta/s}
\end{figure}

The tests for the Green-Kubo approach that have been performed in §\ref{Setting up the Green-Kubo method} now allow us to tackle the complete physical case for the evaluation of $\eta$, in which we consider a mixture of light quarks and gluons, interacting via the proper pQCD matrices which have been introduced in §\ref{Perturbative scattering of massive partons} and extensively written down in Appendix \ref{Appendix A}. The box size and lattice spacing have been set with the values mentioned in §\ref{Setting up the Green-Kubo method} and we averaged the correlator over 45 events. In each simulation the collision of seven particle species ($u,d,s,\Bar{u}.\Bar{d},\bar{s},g$) has been considered, each having relative abundance depending on the Boltzmann distribution and on a proper degeneration factor: for each quark species $3\text{ (colour)}\times 2 \text{ (spin)}$, for the gluons $8\text{ (colour)}\times 2 \text{ (spin)}$. The results obtained in this way are plotted as blue dots in Figure \ref{eta/s}: the error bars represent the statistical error due to the averaging over events, in particular they represent the standard deviation over the mean value.\\
Then, we can evaluate $\eta/s$ in the $1^{\text{st}}$ order Chapman-Enskog approximation. We first considered a bulk of only light quarks ($N_f=3=(u,d,s)$ of equal QPM mass $m_q$ as in \eqref{3.9}) and then the introduction of a charm in the bulk ($N_f=3+1$). In the latter case the introduction of the charm quark has involved only the change in the coefficients of the masses \eqref{3.9} and in the fit to the lattice data for $g(T)$, for which the proper $N_f=3+1$ lQCD data for the energy density have been considered \cite{Borsanyi_2016}. Those curves have been depicted as a full and a dashed red lines in Figure \ref{eta/s}.\\
Unfortunately the formalism for the binary mixture shown in §\ref{binary mixture} does not allow for the interaction among seven different species, therefore the quark-quark cross section has been ‘averaged' as already explained in §\ref{Setting up the Green-Kubo method}. This means that we consider ‘effective quarks' which interact one another with such averaged cross section, and we can apply the two component CE formalism to the effective quark+gluons mixture.\\
What we see is that the CE data agree with the calculations performed within the GK transport approach. From $T\sim 450$ MeV upwards we find that the GK values stand over the red curve: as we highlighted in §\ref{Setting up the Green-Kubo method}, this is likely to be due to the poor approximation that the first order CE provides for the binary mixture (Figure \ref{fig5.1} a) and b)). Instead, in the homogeneous case the first order approximation is already quite satisfactory.

%\begin{figure}[ht]
 %   \centering
%\includegraphics[width=0.8\textwidth]%{etaovers_scaled.eps}
 %   \caption{\textcolor{red}{Cosa facciamo con questa figura? eta/s scalato con il g(T)}.}
%    \label{eta/s_scaled}
%\end{figure}
As extensively shown, the previous results are obtained using a non-isotropic differential cross section. In order to isolate the effect of the anisotropy from the effect due to the magnitude of the total cross section (in the isotropic case, an higher cross section results in a smaller $\eta$), we have performed a CE calculation in which the total cross section $\sigma_{\text{tot}}=\sigma_{\text{tot}}(\sqrt{s},T)$ is kept the same, but instead we choose the differential cross section as $d\sigma/d\Omega=\sigma_{\text{tot}}/4\pi$. By doing so, we basically simulate a system in which the interaction is isotropic, instead of being governed by the pQCD matrices in Appendix \ref{Appendix A}, but whose total cross section is unchanged.\\
The results obtained using the above method are depicted as pink empty circles in Figure \ref{eta/s}: what we deduce is that the high values of $\eta/s$ are due to the significant anisotropy of the scattering matrices. Indeed, in the CE formalism (cf. §\ref{CE homogeneous gas} and §\ref{binary mixture}) we see that the differential cross section appears in the omega integrals as:
$$\omega^{(s)}\propto\int_0^\pi d(\cos\theta)\frac{d\sigma}{d\Omega}\cdot (1-\cos^s\theta),$$
and when $s=2$ the above term is proportional to the \textit{transport cross section} $\sigma_{\text{tr}}$. Such quantity describes the amount of momentum which is exchanged in each collision, and it is sensitive to high-angle contributions. What we find is that the results for $\eta/s$ in which we have considered an ‘equivalent' isotropic cross section assume much lower values with respect to the realistic case. Moreover, in this case $\eta/s\approx 1/4\pi$ almost independently on the temperature for $T\gtrsim T_c$.\\
These isotropic data are in agreement with the results from \cite{Ozvenchuk_2013}. In this work from the PHSD group the collisions are indeed assumed isotropic and the $\eta$ is evaluated using both a Green-Kubo approach (similar to ours) and a Relaxation-Time Approximation (RTA) expression for the collision integral.\\

\begin{figure}[ht]
    \centering
\includegraphics[width=0.85\textwidth]{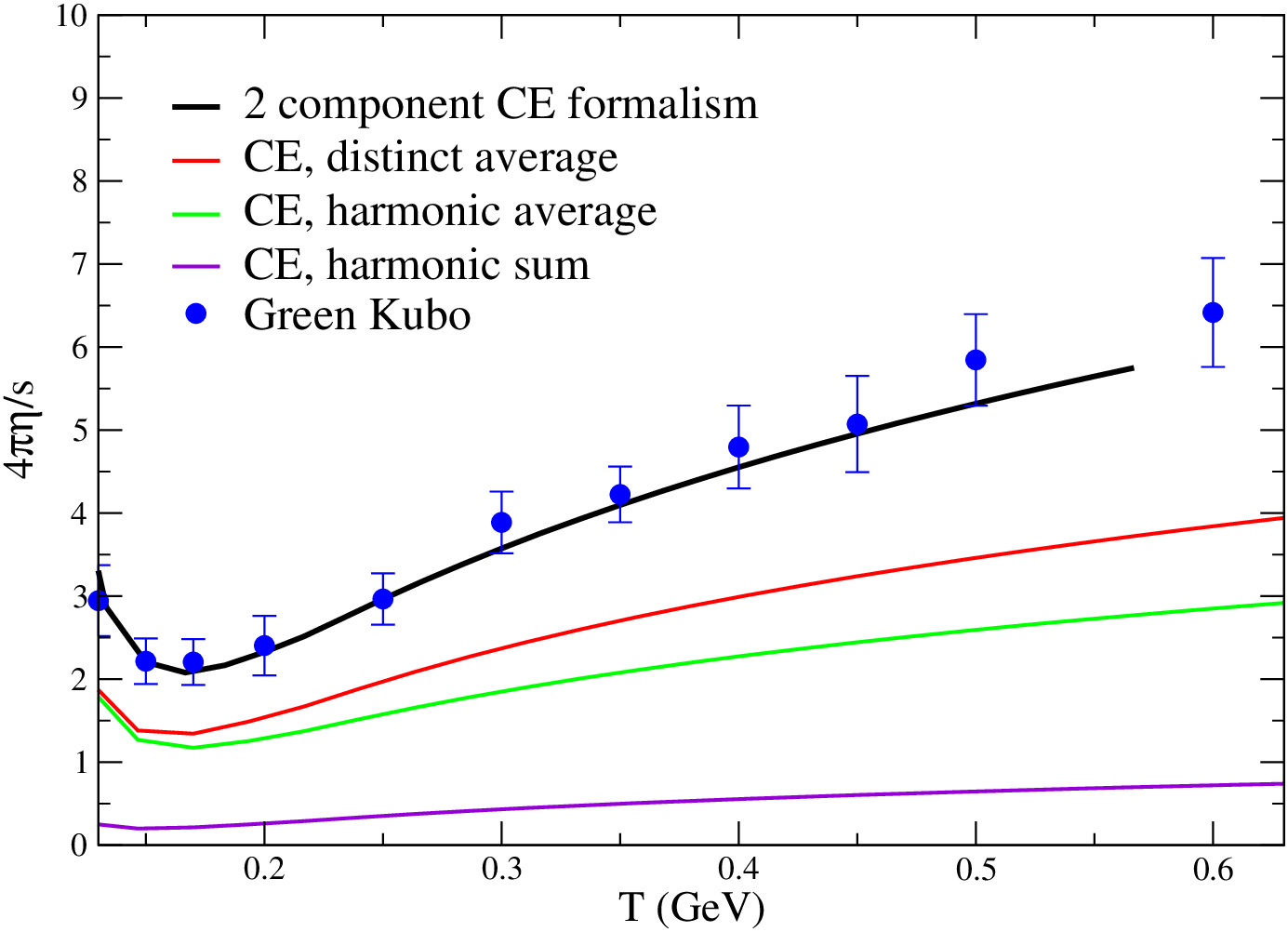}
    \caption{The ratio shear viscosity over entropy density $4\pi \eta/s$ in the two component CE approximation, compared to different ways of ‘averaging' the viscosities of the single homogeneous quark and gluon gases (see text for details).}
    \label{eta/s_averaged}
\end{figure}

As a final point, we want to investigate whether the development of the two component Chapman-Enskog formalism really represents an improvement with respect to more \textit{naive} approaches to the problem. In particular, in Figure \ref{eta/s_averaged} we show our result (in black) compared to different ways of performing the average of $\eta$ of the single quark and gluon gases. The entropy density is kept the same. Specifically, the red line represents the result obtained by averaging the numerator and the denominator of equation \eqref{4.9} separately over quarks and gluons:
$$\left.\eta\right|_{\text{red}}=\frac{T}{10}\frac{d_ge^{-m_g/T} \gamma_{0,g}^2+d_q e^{-m_q/T} \gamma_{0,q}^2}{d_ge^{-m_g/T} c_{00,g}+d_q e^{-m_q/T} c_{00,q}}.$$
In green we instead show the harmonic mean, i.e. the result from the average (with the same weights) of the inverses of the single component viscosities calculated as in \eqref{4.9}:
$$\left.\eta\right|_{\text{green}}=\left(\frac{d_ge^{-m_g/T} \eta_g^{-1}+d_q e^{-m_q/T} \eta_q^{-1}}{d_ge^{-m_g/T}+d_qe^{-m_q/T}}\right)^{-1}.$$
Finally, in violet we show the result obtained by summing the inverse viscosities, without neither any combinatorical nor statistical weight:
$$\left.\eta\right|_{\text{violet}}=\left(\eta_g^{-1}+\eta_q^{-1}\right)^{-1}.$$
All the results are in the $N_f=3$ case.\\
The physical motivation for the latter two approaches lies in the fact that the mean free time for particles in a binary mixture can be \textit{naively} obtained as an harmonic sum of the mean free times of the single components, and the ratio $\eta/s$ is indeed proportional to such time. Anyway, what we see is that only the ‘genuine' two component result is able to reproduce the results from the Green-Kubo transport code (the blue dots in Figure \ref{eta/s_averaged}).

\section{Spatial diffusion coefficient}
\begin{figure}[ht]
    \centering
\includegraphics[width=0.8\textwidth]{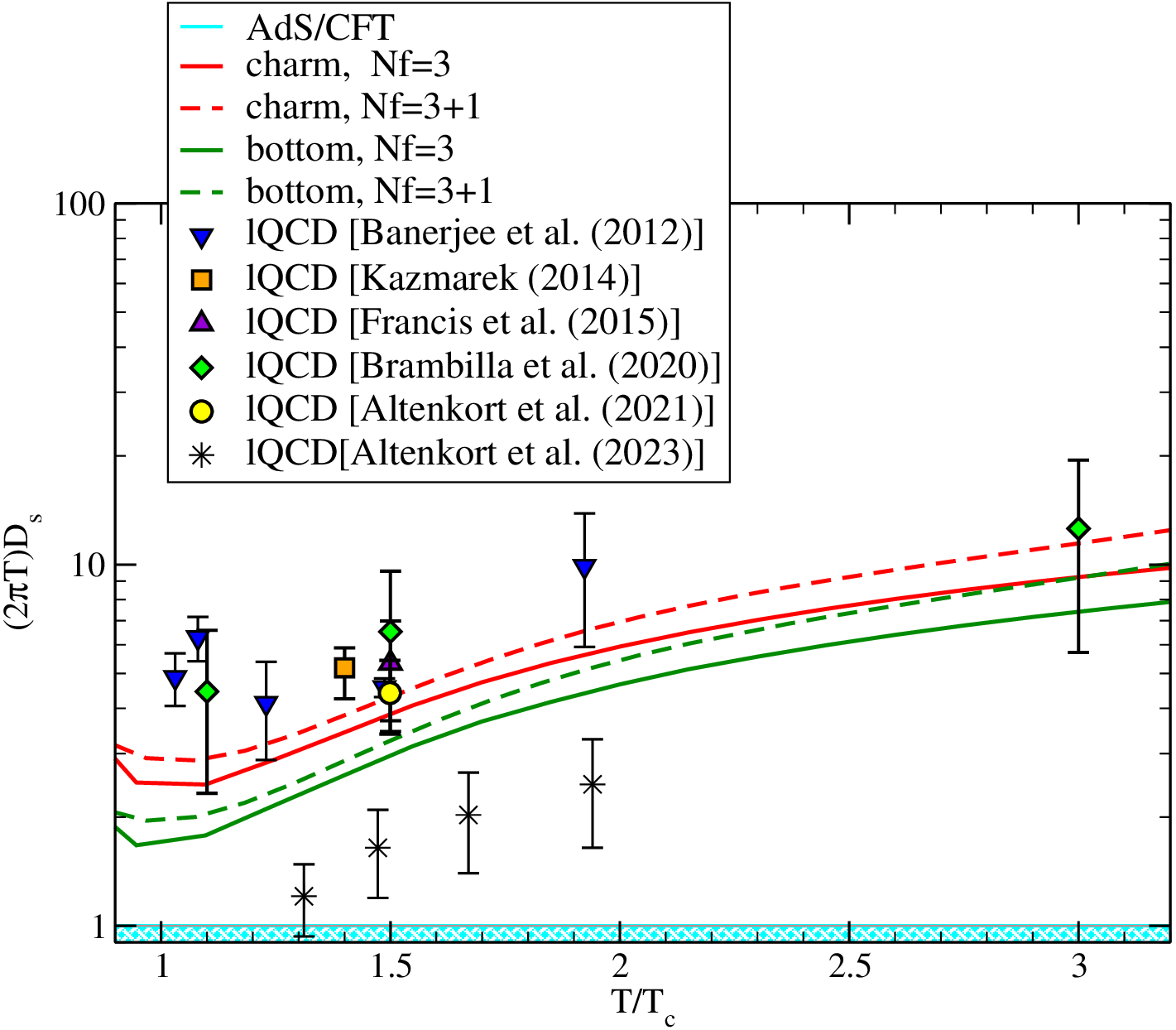}
    \caption{The spatial diffusion coefficient $D_s$ obtained in our approach, considering the diffusion of both a charm and a bottom, in either a light quark or a light quark+charm bulk. We also show the AdS/CFT lower bound $2\pi T D_s\sim 1$ \cite{Gubser_2007} and data from lQCD, taken from \cite{Banerjee_2012} (blue downward triangles), \cite{Kaczmarek_2014} (orange squares), \cite{Francis_2015} (purple upward triangles), \cite{Brambilla_2020} (green losanges), \cite{Altenkort_2021} (yellow circles) and \cite{Altenkort:2023eav} (black stars).}
    \label{fig_Ds}
\end{figure}

In Figure \ref{fig_Ds} we show our results for the temperature dependence of the dimensionless spatial diffusion coefficient $2\pi TD_s$ as given in \eqref{Ds} and calculated by using \eqref{c} and \eqref{4.7} for the drag coefficient. More specifically, we show the results obtained for the diffusion of both a charm quark (of mass $1.3$ GeV) and a bottom quark (of mass $4.2$ GeV). As far as the bulk is concerned, we considered both the $N_f=3$ and the $N_f=3+1$ (see the previous section) cases.\\

\begin{figure}[ht]
    \centering
    \includegraphics[width=0.7\textwidth]{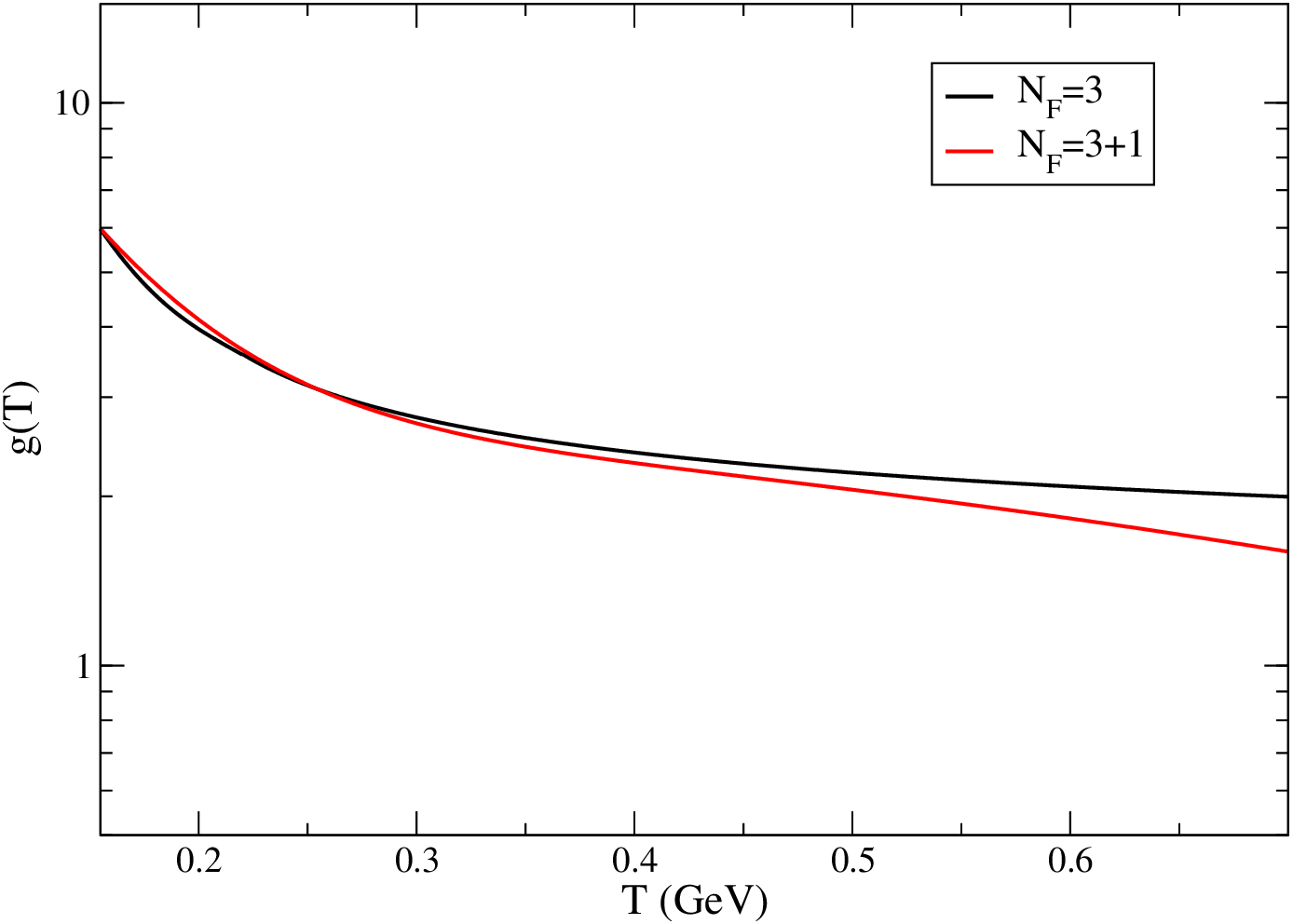}
    \caption{The coupling constants in the QPM for the $N_f=3$ and the $N_f=3+1$ cases.}
    \label{fig_coupling_constants}
\end{figure}

After an initial decrease for $T$, the curves show an absolute minimum for $T\sim T_c$ (here we considered $T_c= 155$ MeV) and then a significant increase for higher temperatures. The green curves obtained for the bottom quark are significantly lower than the corresponding ones for the charm quark, reflecting the inverse dependence with respect to $M_Q$ in \eqref{Ds}. Moreover, each curve obtained with a $N_f=3+1$ bulk stands higher than the corresponding with $N_f=3$. This may be due to the different coupling constants that have to be considered in each case. Indeed, in Figure \ref{fig_coupling_constants} we observe that at high temperatures the $g(T)$ extracted for the case $N_f=3+1$ is smaller than in the case $N_f=3$: this implies that for the case $N_f=3+1$ we estimate a smaller drag coefficient, i.e. a larger $D_s$.

Along with our predictions, in Figure \ref{fig_Ds} we show data obtained in a lattice QCD approach. In particular, it is worthwhile noting that the data from \cite{Banerjee_2012,Kaczmarek_2014,Francis_2015,Brambilla_2020,Altenkort_2021} are obtained in the \textit{quenched approximation}, i.e. by not dynamically including fermions. Instead, the data from \cite{Altenkort:2023eav} (black stars in Figure \ref{fig_Ds}) do not rely on this approximation, however they assume an infinite mass for the heavy quark. From these considerations we deduce that the realistic values for $2\pi T D_s$ have to lie between those two sets of data, which is indeed what we find in our work.
%\begin{figure}[ht]
 %   \centering
%\includegraphics[width=0.8\textwidth]{Ds_scaled.eps}
%    \caption{\textcolor{red}{Cosa facciamo con questa figura? Ds scalato con il g(T)}.}
%    \label{fig_Ds_scaled}
%\end{figure}

\section{The ratio $(2\pi TD_s)/(4\pi \eta/s)$}

In this section we move on to the study of the temperature dependence of the quantity $(2\pi T D_s)/(4\pi \eta/s)$: such a ratio has been proposed as a quantity to better distinguish among strongly- and weakly-coupled media \cite{Rapp_2010,Liu_2020,Min_2022}.\\
In the literature there have been identified two limit cases for such quantity, which will serve their purpose as a comparison with our results \cite{Liu_2020}. In particular, in the strongly coupled limit the correspondence between an anti de Sitter space and a Conformal Field Theory living in its boundary \cite{Maldacena_1999} has been used to estimate the transport coefficients of our interest. The spatial diffusion coefficient was found as $D_s\approx 1/(2\pi T)$ \cite{Gubser_2007,Gubser_2006,Casalderrey_Solana_2006}, whereas the shear viscosity is conjectured to reach a universal strongly-coupled limit of $\eta/s=1/4\pi$ \cite{Kovtun_2005,Buchel_2005}. This implies that in a strongly-coupled system we have:
\begin{equation}
    (2\pi T D_s)\approx 1\cdot (4\pi \eta/s).
    \label{5.a}
\end{equation}
On the other hand, for a weakly-coupled massless gas, the viscosity can be evaluated in a classical kinetic theory as \cite{Danielewicz_1985}:
\begin{equation}
    \eta/s \approx \left(\frac{4}{15}n\langle p \rangle \lambda_{\text{tr}}\right)/s\approx \frac15 T \lambda_{\text{tr}}.
    \label{5.2}
\end{equation}
In the above calculation we have used $n\langle p \rangle =\epsilon$ ($n$ being the particle density) and $Ts=\epsilon+P=4\epsilon/3$: these two relations are valid only in the limit of massless quarks and gluons. Now, we use a momentum transfer mean free path as \cite{Liu_2020}:
\begin{equation}\lambda_{\text{tr}}\approx \tau_{\text{tr}}\approx \tau_Q M_Q/T=D_s,
\label{5.3}
\end{equation}
where $\tau_{\text{tr}}$ and $\tau_Q$ are the light quark mean free time and the heavy quark relaxation time, respectively (once again, in \eqref{5.3} we assumed $m_{q,g}=0$). By doing so, assuming that both \eqref{5.2} and \eqref{5.3} are valid, one obtains the relation:
\begin{equation}
    (2\pi T D_s)\approx 5/2\cdot (4\pi \eta/s).
    \label{5.b}
\end{equation}
for a weakly-coupled system.\\

\begin{figure}[ht]
    \centering
    \includegraphics[width=0.8\textwidth]{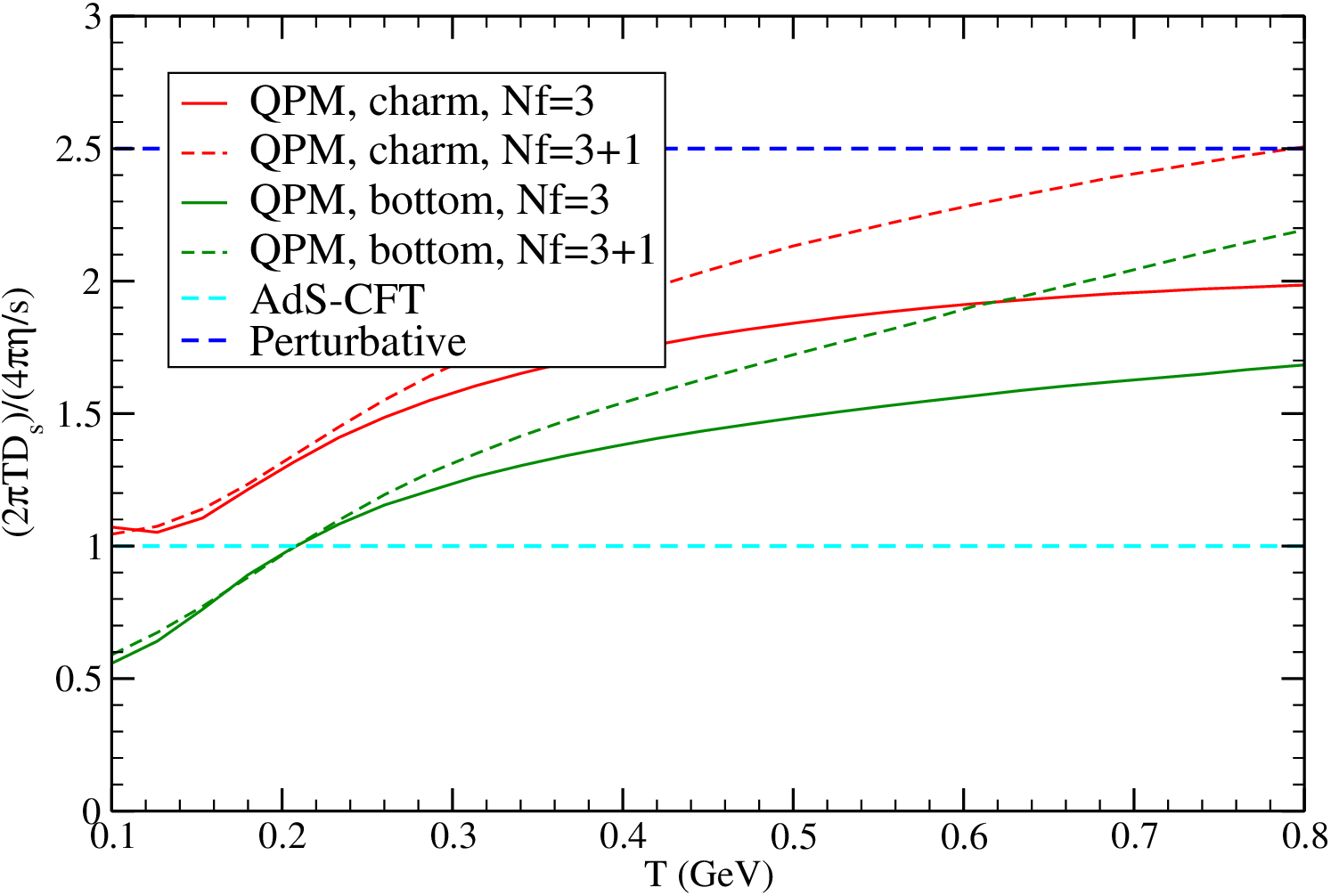}
    \caption{The behaviour of the ratio $(2\pi T D_s)/(4\pi \eta/s)$ with respect to temperature. We also highlight the asymptotic lower boundary from AdS-CFT \cite{Gubser_2007} and the perturbative regime from \cite{Liu_2020}.}
    \label{ratio}
\end{figure}

%\begin{figure}[ht]
 %   \centering
  %  \includegraphics[width=0.8\textwidth]{ratio_scaled.eps}
  %  \caption{\textcolor{red}{Cosa facciamo con questa figura? Ratio scalato con il g(T).}}
  %  \label{ratio_scaled}
%\end{figure}

We show our results (which, however, rely on none of the above approximations) in Figure \ref{ratio}, together with such asymptotic estimates. As we see, in the QPM approach the ratio $(2\pi T D_s)/(4\pi \eta/s)$ is an increasing function of temperature. We observe that for charm quarks at $T\to T_c$ the ratios $(2\pi T D_s)/(4\pi \eta/s)$ of both QPM with $N_f=3$ and $N_f=3+1$ are close to the conjectured limit from the AdS-CFT calculation. This is consistent with the fact that, as shown in Figure \ref{fig_coupling_constants}, the coupling $g(T)$ increases at low temperatures. Note that in the low temperature limit the results for $g(T)$ in both cases show a similar behaviour. On the other hand, at higher temperatures the coupling extracted from QPM is predicted as a decreasing function of $T$: this manifests itself in a larger value for the ratio $(2\pi T D_s)/(4\pi \eta/s)$, closer to the pQCD limit deduced from kinetic theory. Moreover, a comparison between solid and dashed lines shows that such ratio is a quantity which is sensitive to the coupling of the system.\\
As far as the heavy quark mass is concerned, we see that for bottom quarks the ratio is lower than the corresponding results for the charm quark. Indeed, in this case at low temperatures the ratio is even smaller than the estimation coming from the AdS-CFT conjectured limit. At high temperatures, since the quantity $M_Q/T$ gets closer between the charm and bottom cases, the results for the ratio tend to become similar and both curves approach the same pQCD limit.\\

\begin{figure}[ht]
    \centering
    \includegraphics[width=0.8\textwidth]{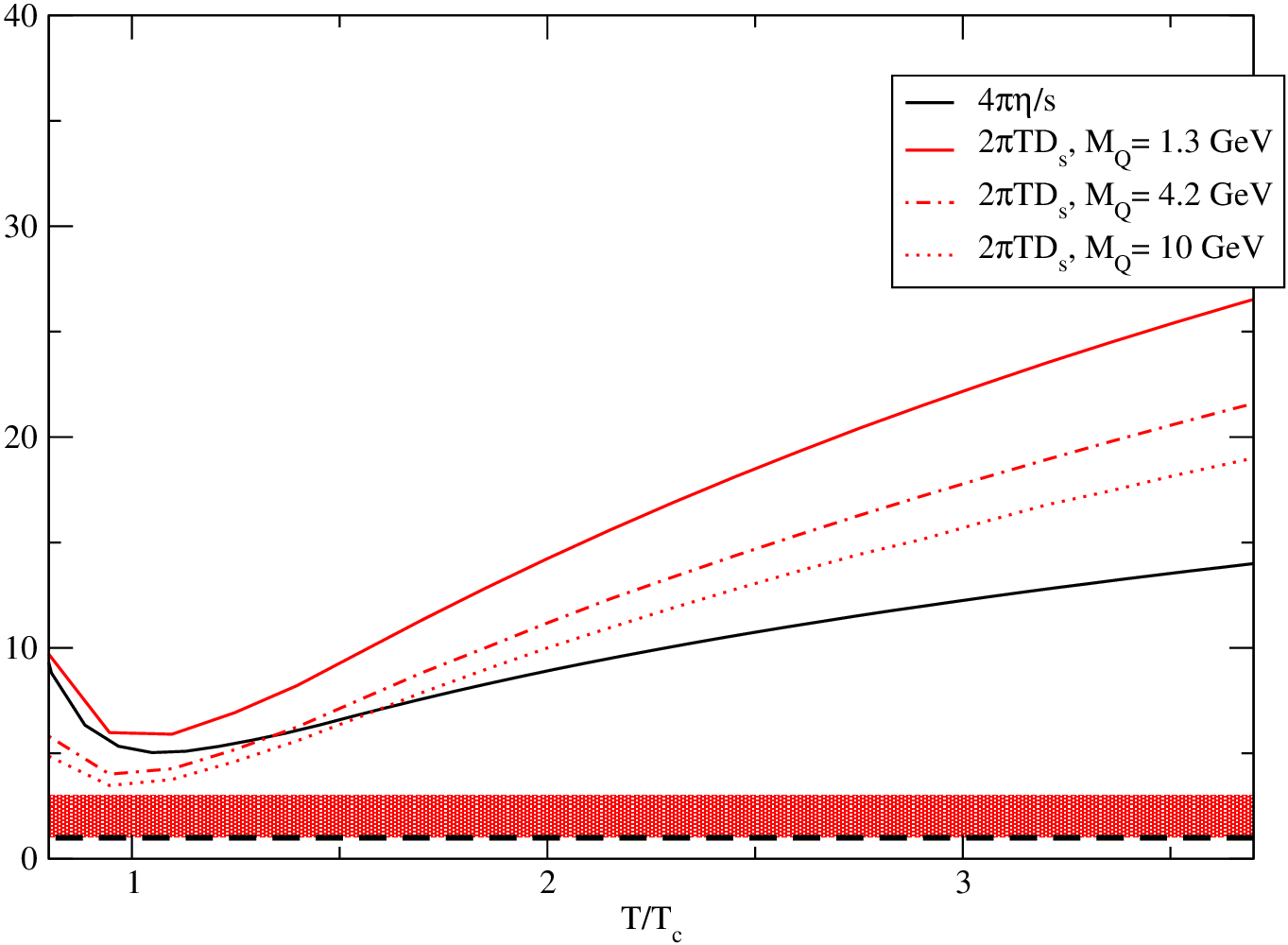}
    \caption{The temperature behaviour (in terms of the scaled temperature $T/T_c$) of the quantities $4\pi \eta/s$ (black) and $2\pi T D_s$  (red, for different masses of the heavy quark).}
    \label{fig:5.6}
\end{figure}

\begin{figure}[ht]
    \centering
    \includegraphics[width=0.8\textwidth]{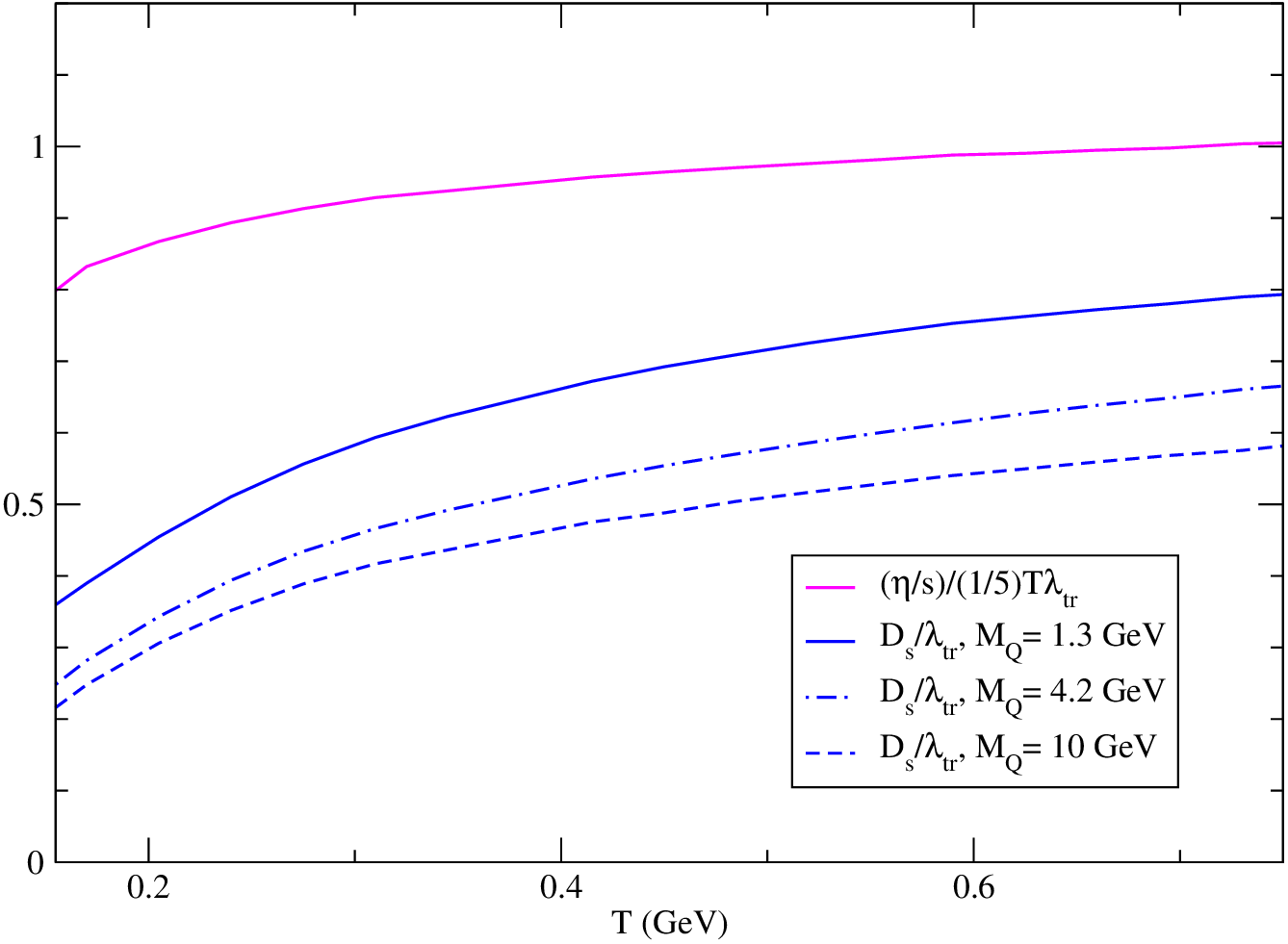}
    \caption{The temperature dependence of $(\eta/s)/(1/5)T\lambda_{\text{tr}}$ (pink) and of $D_s/\lambda_{\tr}$ (blue, for different masses of the heavy quark).}
    \label{fig:5.7}
\end{figure}

In order to investigate more on the validity of the aforementioned perturbative and AdS/CFT limits, in Figure \ref{fig:5.6} we show the behaviour in terms of $T/T_c$ (being $T_c=155$ MeV) of the numerator $2\pi T D_s$ and of the denominator $4\pi \eta/s$ separately. The $D_s$ has been evaluated for both a charm quark, a bottom quark and an hypothetical heavier quark of mass $M_Q=10$ GeV. Interestingly we see that for the charm quark $2\pi T D_s \sim 4\pi \eta/s$ around $T_c$, however for larger $M_Q$ the values of $2\pi T D_s$ get lower, which reflects the fact that the green curves in Figure \ref{ratio} fall lower than the red ones, falling even below the AdS/CFT lower boundary.\\
On the other hand, in order to study the perturbative limit $(2\pi T D_s)\approx 2.5\cdot (4\pi \eta/s)$, we separately consider the approximations \eqref{5.2} and \eqref{5.3} by plotting the temperature dependence of the quantities $(\eta/s)/$ $(1/5)T\lambda_{\text{tr}}$ and $D_s/\lambda_{\tr}$ in Figure \ref{fig:5.7}. The momentum transfer mean free path $\lambda_{\text{tr}}$ has been evaluated as in \cite{Danielewicz_1985}.
In order for \eqref{5.2} and \eqref{5.3} to hold, we expect both ratios to be close to 1 for high temperatures. What we see is that indeed \eqref{5.2} is quite well satisfied for temperatures $T\gtrsim 0.6$ GeV. On the other hand, the ratio $D_s/\lambda_{\text{tr}}$ in the QPM is quite far from 1 even at high temperatures: the closest value comes from the charm quark, whereas for increasing heavy quark masses the high temperature limit gets closer to 0.5. What we find is that, indeed, the weak point of the perturbative estimate lies in the approximation $D_s\sim \lambda_{\text{tr}}$, which for large $M_Q$ is off by around a factor 2.\\

We deduce that the relation $D_s\approx \lambda_{\text{tr}}$ appears to be strongly violated at all temperatures. Furthermore, our approach is applied also to the $T\sim T_c$ region, in which we see that the ratio $(2\pi T D_s)/(4\pi \eta/s)$ can get values quite lower  than the conjectured value in the strong coupling limit. Again, the reason can be attributed to the fact that $D_s/\lambda_{\text{tr}}$ significantly decreases for $T\to T_c$.\\
Our analysis has shown that, in a kinetic theory approach coupled to QPM, the ratio $(2\pi T D_s)/(4\pi \eta/s)$ is an increasing function of $T$: such an increase can be associated to the transition from a quite strong coupling to a weakly coupling regime. However, both the lower conjectured limit from AdS/CFT strong coupling and the weakly coupled limit can be violated, the former quite significantly, the latter even only mildly. The current phenomenological analysis favours, indeed, values of $4\pi \eta/s \approx 2$ and $2\pi T D_s\approx 1-2$ \cite{Gubser_2007}, which translate onto a ratio $(2\pi T D_s)/(4\pi \eta/s)$ with values lower than 1, which are reflected in our approach. It can be envisaged that upcoming advances in both lattice QCD and phenomenological studies, along with the upcoming new experimental data on heavy flavours, may put more quantitative constraints to these quantities.

\chapter*{\iflanguage{italian}{Conclusioni}{Conclusions}}
\addcontentsline{toc}{chapter}{\iflanguage{italian}{Conclusioni}{Conclusions}}
\markboth{\MakeUppercase{Conclusions}}{\MakeUppercase{Conclusions}}

The comprehension of the deconfined phase of quarks and gluons plays a key role in the understanding of the mechanisms of the Quantum ChromoDynamics. Although the natural starting point for the study of such high temperature region is perturbative QCD, there are evidences that the QGP is a strongly interacting system, with the result that the strong coupling constant assumes values significantly larger than the perturbative range at high energy. A perturbative study is not therefore fully appropriate for the description of such plasma state and we need to take into account non-perturbative dynamics. One way to do so is via the application of a Quasi-Particle Model (QPM), in which the interaction is encoded in the partonic T-dependent quasi-particle masses: the strong coupling $g(T)$ is then obtained by fitting the lQCD energy density $\epsilon$ of the system.\\
Moreover, the \textit{heavy quarks} (charm and bottom) represent an excellent probe
for the study of QGP due to their large mass, leading to a formation time for these particles in nucleus-nucleus collisions smaller than the formation time of QGP itself. These particles do
not reach thermodynamic equilibrium with the dense medium, so their evolution is studied by means of microscopic equations, such as the relativistic Boltzmann transport equation. In the approximation of small momentum transferred, their dynamics is often assimilated to a Brownian-type motion and described by the Fokker-Planck equation. In this approach, the
effect of the interaction between heavy quarks and bulk partons is described in terms
of the transport coefficients. \\

In order to study the interaction among quasi-particles of the bulk and between quasi-particles and heavy quarks, in the first part of this thesis we have developed a scattering matrix formalism which takes into account a non-zero mass of the gauge bosons (gluons). In particular, we have evaluated the differential cross section $d\sigma/d\Omega$ for the various processes under consideration ($qq\to qq$, $qg\to qg$, $gg\to gg$) by taking into account the relevant contributions at tree level and considering the quarks and gluons within the QPM.\\
Once such a framework has been established, we have illustrated the methods which have been used to derive the observables of our interest. Namely, for the shear viscosity $\eta$ we employed both the Chapman-Enskog approximation (based on a microscopic implementation of a gradient expansion) and the Green-Kubo method (based on the relativistic Boltzmann equation), whereas the Fokker-Planck equation has allowed the derivation of the spatial diffusion coefficient $D_s$ of a heavy quark in the medium. The relevant novelty of this thesis in those topics lies in the development of the two-component Chapman-Enskog formalism within a QPM and its comparison with the Green-Kubo method, for which a relativistic transport code has been set up.\\

%We have evaluated the ratio $\eta/s$ in the CE formalism, being $s$ the entropy density, and shown that these results are significantly greater than the ones coming from lQCD and from calculations by other groups (e.g. PHSD). 
We have seen that our CE results agree with the GK calculations performed employing a simulation code for the relativistic Boltzmann equation. This indicates that the 2-component formula within the CE approach is able to correctly account for the $\eta/s$ of the fluid. Furthermore, we have shown that the conjectured formulas that try to estimate $\eta/s$ as averages over the single fluid components can largely underestimate the $\eta/s$ of the 2-component fluid. It may be useful to explore the multi-component (3+) Chapman-Enskog formalism, which will allow us to explore each interaction separately without the need of any ‘average' procedure for the quark cross sections: such results may then once again be compared to the Green-Kubo results. This may be a subject for future work. \\
Moreover, the spatial diffusion coefficient $D_s$ evaluated from the Fokker-Planck equation shows a relevant temperature dependence, approaching its minimal value for $T\sim T_c$ and showing significant agreement with the lattice data.\\
%Such high values for the ratio $\eta/s$ are likely to be due to the fundamental anisotropy of our cross sections: indeed, we have checked that by fixing the total cross section and instead considering the ‘equivalent' isotropic system, the results are sensibly lower.\\

Finally, we have studied the ratio $(2 \pi T D_s)/(4 \pi \eta/s)$, which has been suggested to be used as a quantity to better distinguish among strongly- and weakly-coupled media. In particular, it has been conjectured to range between 1 and 2.5 while going from strong to weak coupling. We observe that within the QPM, this ratio for the charm quarks indicates the presence of a strong coupling regime for low temperatures and a weaker coupling regime for higher temperatures, the latter being consistent with the perturbative QCD limit. This indeed shows that the Quasi-Particle Model is able to describe the transition from a strong coupling system to a weakly coupled one.\\
Nonetheless, a more thorough analysis shows that the approximations which have been used to describe the strong/weak coupling regimes appear questionable. Indeed one sees that $D_s\sim \lambda_{\text{tr}}$ is not enough accurate neither in the strong nor in the weak coupling regimes at very high temperatures. Moreover, the approximation gets increasingly worse for larger heavy quark masses (namely, the bottom quark). The direct connection of $D_s$ to the $\lambda_{\text{tr}}$ of the bulk matter is in fact quite extreme even in a realistic non-conformal strong coupling regime: it can therefore occur that the lower bound $(2 \pi T D_s)/(4 \pi \eta/s)>1$ is violated. Indeed, the current determination of the spatial diffusion coefficient gives $2\pi T D_s\sim 1-3$ within the most recent calculations of lattice QCD, giving a value very close to 1 at $T \simeq T_c$, while estimating $4\pi \eta/s\sim 1-2.5$. This means that, within the aforementioned uncertainties, $(2 \pi T D_s)/(4 \pi \eta/s)$ can be even lower than 1, as suggested by the present study within the QPM approach.\\

%This can be seen by noting that for large heavy quark masses the temperature cannot be assumed to be the only energy scale in our problem, even $T\sim 1$ GeV. In other words, in those cases the ratio $M_Q/T$ maintains a significant dependence on $M_Q$ for increasing $T$ and we will not reach a perturbative regime anytime soon.\\

The comparison between experimental data and simulations at RHIC and LHC energies will provide more powerful constraints for the corresponding transport coefficients. In particular, the upgrades of ALICE and CMS detectors will allow to access the low momentum region with a significantly improved precision: it is therefore likely that, in the near future, we will be able to explore also the momentum dependence of the interaction among the heavy quarks and the partons in the bulk. In this sense, the work of this thesis may be extended within a momentum-dependent QPM.\\
Finally, a possible development may consist in the calculation of higher order pQCD scattering matrices, i.e. in the inclusion of loops in the Feynman diagrams for the processes of our interest.
%thermal resummations??

\appendix
\chapter{Massive pQCD scattering matrices}
\label{Appendix A}

In this Appendix we give more details on the calculation of the matrix elements used to evaluate the cross sections for massive partons. In order to fix the notation, let us remind that, by calling $(k_i,p_i)$ the initial four momenta of the particles and $(k_f,p_f)$ the final four momenta, the three Mandelstam variables are given by:
$$s=(k_i+p_i)^2=(k_f+p_f)^2,~~t=(k_i-k_f)^2=(p_i-p_f)^2,$$
$$u=(k_i-p_f)^2=(p_i-k_f)^2.$$
Moreover, the generators of SU(3) associated with QCD are the Gell-Mann matrices divided by two, i.e. $T^a=\lambda^a/2$ being $a$ the gluon colour index $a=1,\dots, 8$ \cite{Gell-Mann_1962}. The Lie algebra which the generators $T^a$ have to obey is given by the commutation relation:
$$[T^a,T^b]=if^{abc}T^c,$$
where $f^{abc}$ are the SU(3) structure constants. The rules for the colour sums which we are going to develop are given in detail in \cite{MacFarlane_1968}.\\
Furthermore, the Dirac gamma matrices will be denoted by $\gamma^\mu$, $u$ and $v$ will indicate the Dirac spinors for particles and anti-particles, respectively. Finally, $\epsilon$ will denote the polarization vector for gluons.\\

We are going to proceed by listing all the matrix elements $\mathcal{M}$ of our interest, as derived from the Feynman rules applied to the proper diagrams. The final analytical expressions for $|\mathcal{M}|^2$ (properly averaged over initial - and summed over final - spins and colours) have been evaluated using \textsc{feyncalc} \cite{Mertig_1991,Shtabovenko_2016,Shtabovenko_2020}.\\

The same matrices have been used not only for the evaluation of $\eta$ but also for the $D_s$, by considering the expressions for $gq\to gq$,  $qq'\to qq'$ and then substituting the heavy quark mass. However, since the heavy quark under consideration may collide with partons having different spin, flavour and colour, we have multiplied the resulting $gQ\to gQ$ squared scattering matrix by $8 \text{ (colours)}\times 2 \text{ (polarizations)}$, and the $qQ\to qQ$ squared scattering matrix by $3\text{ (colours)}\times 2 \text{ (spins)} \times 3\text{ (flavours)}\times 2 \text{ (antiparticles)}$. This must not done for the evaluation of $\eta$, i.e. when considering the collisions among the bulk partons, since these factors are already included in the relative abundance of the quark-gluon mixture.\\
Moreover, note that in the $Qq\to Qq$ process only the direct diagram is involved, since the heavy quark and the light quark from the bulk will always be different. In other words, in this case the annihilation/exchange diagrams do not have to be considered.

\section{Quark-quark scattering}
When dealing with quark-quark scattering, depending on the flavours we may have the presence of not only the direct ($t$-channel) diagram, but also the exchange or the annihilation diagram ($u$- and $s$-channels, respectively). The diagrams involved in the quark-quark scattering are depicted in Figure \ref{qq_FD}.
\begin{fmffile}{quarkquark}
\begin{figure}[H]
\begin{minipage}[ht]{0.49\textwidth}
\centering
\hspace*{8mm}\begin{fmfgraph*}(120,80)
    \fmfleft{i1,i2}
    \fmfright{o1,o2}
    \fmf{gluon, width=5}{v1,v2}
    \fmf{fermion, width=5}{i2,v1}
    \fmf{fermion, width=5}{v1,o2}
    \fmf{fermion, width=5}{i1,v2}
    \fmf{fermion,  width=5}{v2,o1}
    \fmflabel{$q'$}{i1}
    \fmflabel{$q$}{i2}
    \fmflabel{$q'$}{o1}
    \fmflabel{$q$}{o2}
\end{fmfgraph*}
\caption*{t-channel}
\end{minipage}
\begin{minipage}[ht]{0.49\textwidth}
\centering
\hspace*{8mm}\begin{fmfgraph*}(120,80)
    \fmftop{i1,i2}
    \fmfbottom{o1,o2}
    \fmf{fermion, tension=2, width=5}{i1,v2}
    \fmf{gluon, tension=0.3, width=5}{v2,w2}
    \fmf{fermion, tension=2, width=5}{o1,w2}
    \fmf{phantom}{w2,o2}
    \fmf{phantom}{i2,v2}
    \fmffreeze
    \fmf{fermion, width=5}{v2,o2}
    \fmf{fermion, width=5}{w2,i2}
    \fmflabel{$q$}{i1}
    \fmflabel{$q$}{i2}
    \fmflabel{$q$}{o1}
    \fmflabel{$q$}{o2}
\end{fmfgraph*}\caption*{u-channel}
\end{minipage}
\end{figure}
\vspace{1em}
\begin{figure}[H]
\centering
\hspace*{8mm}\begin{fmfgraph*}(120,80)
    \fmftop{i1,i2}
    \fmfbottom{o1,o2}
    \fmf{gluon, width=5}{v1,v2}
    \fmf{fermion, width=5}{i2,v1,o2}
    \fmf{fermion, width=5}{i1,v2,o1}
    \fmflabel{$q$}{i1}
    \fmflabel{$q$}{i2}
    \fmflabel{$\Bar{q}$}{o1}
    \fmflabel{$\Bar{q}$}{o2}
\end{fmfgraph*}
\caption*{s-channel}
\caption{Leading-order Feynman diagrams of the quark-quark scatterings.}
\label{qq_FD}
\end{figure}
\end{fmffile}

By denoting as $\alpha,\beta,\gamma,\delta$ the flavour indexes, $i,j,k,l=1,2,3$ the quark colour indexes and $a,b,c,d,e=1,\dots,8$ the gluon colour indexes, each of the above diagrams carries the following contribution for the $\mathcal{M}$ matrix \cite{Moreau_2019}:

\begin{align*}
i\mathcal{M}_t(q_a^iq_\beta^k&\to q_\delta^j q_\gamma^l)=\delta_{\alpha\delta}\delta_{\beta\gamma}\Bar{u}_\delta^j (k_f)(-ig\gamma^\mu T_{ij}^a)u_\alpha^i(k_i)\cdot\\
&\cdot\left[-i\frac{g_{\mu\nu}-(q_\mu^tq_\nu^t)/M_g^2}{(k_f-k_i)^2-M_g^2}\right]\Bar{u}_\gamma^l(p_f)(-ig\gamma^\nu T_{kl}^a)u_\beta^k(p_i)
\end{align*}

\begin{align*}
i\mathcal{M}_u(q_a^iq_\beta^k&\to q_\delta^j q_\gamma^l)=-\delta_{\alpha\delta}\delta_{\alpha\delta}\delta_{\beta\gamma}\Bar{u}_\delta^j (k_f)(-ig\gamma^\nu T_{kj}^a)u_\beta^k(p_i)\cdot\\
&\cdot\left[-i\frac{g_{\mu\nu}-(q_\mu^uq_\nu^u)/M_g^2}{(p_f-k_i)^2-M_g^2}\right]\Bar{u}_\gamma^l(p_f)(-ig\gamma^\mu T_{il}^a)u_\alpha^i(k_i)
\end{align*}

\begin{align*}
i\mathcal{M}_s(q_a^iq_\beta^k&\to q_\delta^j q_\gamma^l)=-\delta_{\alpha\Bar{\beta}}\delta_{\delta \Bar{\gamma}}\Bar{u}_\delta^j (k_f)(-ig\gamma^\nu T_{lj}^a)v_\gamma^l(p_f)\cdot\\
&\cdot\left[-i\frac{g_{\mu\nu}-(q_\mu^sq_\nu^s)/M_g^2}{(k_i+p_i)^2-M_g^2}\right]\Bar{v}_\beta^k(p_i)(-ig\gamma^\mu T_{ik}^a)u_\alpha^i(k_i)
\end{align*}
where $q^\mu_t=(k_f-k_i)^\mu$, $q^\mu_u=(p_i-k_f)^\mu$ and $q_s^\mu=(k_i+p_i)^\mu$ is the momentum of the exchanged gluon in each case. These matrices are then added up, modulus-squared and averaged (summed) over initial (final) partons, that is:
$$|\mathcal{M}(q_\alpha q_\beta\to q_\delta q_\gamma)|^2=\frac{1}{3\times 2}\frac{1}{3\times 2}\sum_{\text{colour}}\sum_{\text{spin}}|\mathcal{M}_t+\mathcal{M}_u+\mathcal{M}_s|^2$$

The action of the Kronecker deltas in the previous expressions for $\mathcal{M}$ therefore selects the diagrams of our interest depending on the quarks involved. In practice, for quarks of different flavours we are going to consider only the $t$-channel diagram, for $q\Bar{q}\to q\Bar{q}$ we are considering both the $t$- and the $s$- channel diagrams and for $qq\to qq$ we are summing $t$- and $u$- channel diagrams.\\

The final results are:

\begin{align*}
    |\mathcal{M}(qq'\to qq')|^2&=\\
    &\frac{4g^4}{9(M_g^2-t)^2}\cdot\left[2M_q^4+2M_q^2(2M_{q'}^2-s+t-u)+2M_{q'}^4-\right.\\
   &2M_{q'}^2(s-t+u)+\left.s^2+u^2\right],\\
\end{align*}

\begin{align*}
|\mathcal{M}&(qq\to qq)|^2=\\
&\frac{4g^4}{27(M_g^2-t)^2(M_g^2-u)^2}\cdot\left\{8M_q^4[5M_g^4-5M_g^2(t+u)+3(t^2+u^2)-\right.\\
&tu]-4M_q^2(M_g^4(3s+t+u)-M_g^2(3s(t+u)+7t^2-10tu+7u^2)+\\
&3s(t^2+u^2)-3stu+(t+u)(3(t-u)^2+tu))+M_g^4(4s^2+3(t^2+u^2))-\\
&\left.2M_g^2(t+u)(2s^2+3(t^2-tu+u^2))+3s^2t^2-2s^2tu+3s^2u^2+3t^4+3u^4\right\},
\end{align*}

\begin{align*}
|\mathcal{M}&(q\Bar{q}\to q\Bar{q})|^2=\\
&\frac{4g^4}{27(M_g^2-s)^2(M_g^2-t)^2}\cdot\left\{8M_q^4[5M_g^4-5M_g^2(s+t)+3(s^2+t^2)-\right.\\
&st]-4M_q^2(M_g^4(3u+s+t)-M_g^2(3u(s+t)+7s^2-10st+7t^2)+\\
&3u(s^2+t^2)-3stu+(s+t)(3(s-t)^2+st))+M_g^4(4u^2+3(s^2+t^2))-\\
&\left.2M_g^2(s+t)(2u^2+3(s^2-st+t^2))+3u^2s^2-2u^2ts+3u^2t^2+3s^4+3t^4\right\}.
\end{align*}

\newpage

\section{Quark-gluon scattering}
The diagrams involved in the $qg\to qg$ scattering are depicted in Figure \ref{gq_FD}.
\begin{fmffile}{quarkgluon}
\begin{figure}[H]
\begin{minipage}[ht]{0.49\textwidth}
\centering
\hspace*{8mm}\begin{fmfgraph*}(120,80)
    \fmfleft{i1,i2}
    \fmfright{o1,o2}
    \fmf{gluon, width=5}{v1,v2}
    \fmf{gluon, width=5}{i2,v1}
    \fmf{gluon, width=5}{v1,o2}
    \fmf{fermion, width=5}{i1,v2}
    \fmf{fermion, width=5}{v2,o1}
    \fmflabel{$q$}{i1}
    \fmflabel{$g$}{i2}
    \fmflabel{$q$}{o1}
    \fmflabel{$g$}{o2}
\end{fmfgraph*}
\caption*{t-channel}
\end{minipage}
\begin{minipage}[ht]{0.49\textwidth}
\centering
\hspace*{8mm}\begin{fmfgraph*}(120,80)
    \fmfleft{i1,i2}
    \fmfright{o1,o2}
    \fmf{fermion, width=5}{v2,v1}
    \fmf{gluon, width=5}{i2,v1}
    \fmf{fermion, width=5}{v1,o2}
    \fmf{fermion, width=5}{i1,v2}
    \fmf{gluon, width=5}{v2,o1}
    \fmflabel{$q$}{i1}
    \fmflabel{$g$}{i2}
    \fmflabel{$g$}{o1}
    \fmflabel{$q$}{o2}
\end{fmfgraph*}
\caption*{u-channel}
\end{minipage}
\end{figure}
\vspace{1em}
\begin{figure}[H]
\centering
\hspace*{8mm}\begin{fmfgraph*}(120,80)
    \fmftop{i1,i2}
    \fmfbottom{o1,o2}
    \fmf{fermion, width=5}{o1,v1,v2,o2}
    \fmf{gluon, width=5}{i1,v1}
    \fmf{gluon, width=5}{i2,v2}
    \fmflabel{$g$}{i1}
    \fmflabel{$g$}{i2}
    \fmflabel{$q$}{o1}
    \fmflabel{$q$}{o2}
\end{fmfgraph*}
\caption*{s-channel}
\caption{Leading-order Feynman diagrams of the $qg\to qg$ scatterings.}
\label{gq_FD}
\end{figure}
\end{fmffile}
These diagrams correspond to the following expressions for the invariant matrix elements \cite{Moreau_2019}:

\begin{align*}
i\mathcal{M}_t(g^a q^i\to g^b q^j)&=(\epsilon_{b,f}^*)_\nu[-g f^{cab}C^{\lambda \mu\nu}(k_i-k_f,-k_i,k_f)](\epsilon_{a,i})_\mu\cdot\\
&\cdot\left[-i\frac{g_{\lambda \tau}-(q_\lambda^tq_\tau^t)/M_g^2}{(k_f-k_i)^2-M_g^2}\right]\Bar{u}^j(p_f)(-ig\gamma^\tau T_{ij}^c)u^i(p_i),
\end{align*}

\begin{align*}
i\mathcal{M}_u(g^a q^i\to g^b q^j)=\Bar{u}^j(p_f)(&-ig\gamma^\mu T_{kj}^a))(\epsilon_{a,i})_\mu\cdot\\
&\cdot\left[i\frac{\slashed{q}^u+M_q}{u-M_g^2}\right](\epsilon_{b,f}^*)_\nu (-ig\gamma^\nu T_{ik}^b)u^i(p_i),
\end{align*}

\begin{align*}
i\mathcal{M}_s(g^a q^i\to g^b q^j)=\Bar{u}^j(p_f)(&-ig\gamma^\nu T_{lj}^b))(\epsilon_{b,f}^*)_\nu\cdot\\
&\cdot\left[i\frac{\slashed{q}^s+M_q}{s-M_g^2}\right](\epsilon_{a,i})_\mu (-ig\gamma^\mu T_{il}^a)u^i(p_i),
\end{align*}
with the 3-gluon vertex being:\footnote{The three momenta are considered as all entering the vertex.}
$$C^{\lambda\mu\nu}(q_1,q_2,q_3)=[(q_1-q_2)^\nu g^{\lambda\mu}+(q_2-q_3)^\lambda g^{\mu\nu}+(q_3-q_1)^\mu g^{\lambda \nu}].$$
Let us remind that, since we are assuming a massive gluon, the sum over polarizations is performed as:
$$\sum_{\text{pol.}}(\epsilon_i)_\mu (\epsilon_i^*)_{\mu'}=-g_{\mu\mu'}+\frac{(k_i)_\mu (k_i)_{\mu'}}{M_g^2}.$$
Now, we will evaluate:
$$|\mathcal{M}(gq\to gq)|^2=\frac{1}{8\times 2}\frac{1}{3\times 2}\sum_{\text{colour}}\sum_{\text{spin}}|\mathcal{M}_t+\mathcal{M}_u
+\mathcal{M}_s|^2,$$
and he final result is:

\begin{align*}
    |\mathcal{M}(&qg\to qg)|^2=\\
    g^4&\{207M_q^{12}+9M_q^{10}(8M_g^2-21(s+u))-\\
M_q^8&(255M_g^4-8M_g^2(45s+3t+45u)+27s^2+585su+12t^2+27u^2)+\\
9M_q^6&(-8M_g^6+82M_g^4(s+u)-4M_g^2(11s^2+62su+11u^2)+\\
&17s^3+105s^2u+105su^2+17u^3)+\\
M_q^4&(48M_g^8+4M_g^6(43s-6t+43u)+M_g^4(-291s^2+4s(4t-551u)+\\
&12t^2+16tu-291u^2)+4M_g^2(54s^3+s^2(450u-3t)-\\
&2s(t^2+7tu-225u^2)+u(-2t^2-3tu+54u^2))-90s^4-\\
&603s^3u+6s^2t^2-693s^2u^2+28st^2u-603su^3+6t^2u^2-90u^4)+\\
M_q^2&(-48M_g^8(s+u)+4M_g^6(9s^2+2s(3t-68u)+u(6t+9u))+\\
&2M_g^4(8s^3+533s^2u-s(6t^2+32tu-533u^2)-6t^2u-8u^3)+\\
&4M_g^2(9s^4-s^3(t-189u)+s^2u(162u-7t)+su(-8t^2-7tu+189u^2)+\\
&u^3(9u-t))+(s+u)(18s^4+144s^3u+s^2(63u^2-2t^2)-\\
&4s(3t^2u-36u^3)-2t^2u^2+18u^4))+\\
2M_g^8&(8s^2+8su+8u^2)-4M_g^6(9s^3-s^2(t+34u)+2su(4t-17u)+\\
&u^2(9u-t))+M_g^4(18s^4-232s^3u+s^2(-2t^2+16tu-171u^2)+\\
&2su(8t^2+8tu-116u^2)-2t^2u^2+18u^4)+4M_g^2su(27s^3-s^2(t-27u)+\\
&s(27u^2-2t^2)+u(-2t^2-tu+27u^2))+su(-18s^4+s^2(2t^2-45u^2)+\\
&2u^2(t^2-9u^2))\}/[36(M_q^2-s)^2(M_q^2-u)^2(M_g^2-t)^2].
\end{align*}

\section{Gluon-gluon scattering}
The diagrams involved in the $gg\to gg$ scattering are depicted in Figure \ref{gg_FD}.
\begin{fmffile}{gluongluon}
\begin{figure}[H]
\begin{minipage}[ht]{0.49\textwidth}
\centering
\hspace*{8mm}\begin{fmfgraph*}(120,80)
    \fmfleft{i1,i2}
    \fmfright{o1,o2}
    \fmf{gluon, width=5}{v1,v2}
    \fmf{gluon, width=5}{i2,v1}
    \fmf{gluon, width=5}{v1,o2}
    \fmf{gluon, width=5}{i1,v2}
    \fmf{gluon, width=5}{v2,o1}
    \fmflabel{$g$}{i1}
    \fmflabel{$g$}{i2}
    \fmflabel{$g$}{o1}
    \fmflabel{$g$}{o2}
\end{fmfgraph*}\caption*{t-channel}
\end{minipage}
\begin{minipage}[ht]{0.49\textwidth}
\centering
\hspace*{8mm}\begin{fmfgraph*}(120,80)
    \fmftop{i1,i2}
    \fmfbottom{o1,o2}
    \fmf{gluon, tension=2, width=5}{i1,v2}
    \fmf{gluon, tension=0.3, width=5}{v2,w2}
    \fmf{gluon, tension=2, width=5}{o1,w2}
    \fmf{phantom}{w2,o2}
    \fmf{phantom}{i2,v2}
    \fmffreeze
    \fmf{gluon, width=5}{v2,o2}
    \fmf{gluon, width=5}{w2,i2}
    \fmflabel{$g$}{i1}
    \fmflabel{$g$}{i2}
    \fmflabel{$g$}{o1}
    \fmflabel{$g$}{o2}
\end{fmfgraph*}
\caption*{u-channel}
\end{minipage}
\end{figure}
\begin{figure}[H]
\begin{minipage}[ht]{0.49\textwidth}
\centering
\hspace*{8mm}\begin{fmfgraph*}(120,80)
    \fmftop{i1,i2}
    \fmfbottom{o1,o2}
    \fmf{gluon, width=5}{v1,v2}
    \fmf{gluon, width=5}{i2,v1}
    \fmf{gluon, width=5}{v1,o2}
    \fmf{gluon, width=5}{i1,v2}
    \fmf{gluon, width=5}{v2,o1}
    \fmflabel{$g$}{i1}
    \fmflabel{$g$}{i2}
    \fmflabel{$g$}{o1}
    \fmflabel{$g$}{o2}
\end{fmfgraph*}
\caption*{s-channel}
\end{minipage}
\begin{minipage}[ht]{0.49\textwidth}
\centering
\hspace*{8mm}\begin{fmfgraph*}(120,80)
    \fmfleft{i1,i2}
    \fmfright{o1,o2}
    \fmf{gluon, width=5}{i2,v1}
    \fmf{gluon, width=5}{v1,o2}
    \fmf{gluon, width=5}{i1,v1}
    \fmf{gluon, width=5}{v1,o1}
    \fmflabel{$g$}{i1}
    \fmflabel{$g$}{i2}
    \fmflabel{$g$}{o1}
    \fmflabel{$g$}{o2}
\end{fmfgraph*}
\caption*{4-point}
\end{minipage}
\caption{Leading-order Feynman diagrams of the $gg\to gg$ scatterings.}
\label{gg_FD}
\end{figure}
\end{fmffile}
Such diagrams lead to the following expressions for the invariant matrix elements \cite{Moreau_2019}:\footnote{In the expression for $\mathcal{M}_4$, the minus sign in the term $-f^{ade}f^{cbe}$ appears as a plus in \cite{Moreau_2019}, but if this were a plus not even the usual pQCD $gg\to gg$ calculation (with massless gluons) would yield the result for $d\sigma/d\Omega$ found in literature (e.g. in \cite{Ellis_stirling_webber_1996}).}

\begin{align*}
i&\mathcal{M}_t(g^a g^b \to g^c g^d)=(\epsilon_{d,4}^*)_\sigma[-g f^{ead}C^{\tau\lambda\sigma}(q_1-q_4,-q_1,q_4)](\epsilon_{a,1})_\lambda\cdot\\
&\cdot\left[-i\frac{g_{\tau \tau'}-(q_\tau^tq_{\tau'}^t)/M_g^2}{(q_4-q_1)^2-M_g^2}\right](\epsilon_{c,3}^*)_\nu [-g f^{ecb}C^{\tau'\nu\mu}(-q_3+q_2,q_3,-q_2)](\epsilon_{b,2})_\mu,
\end{align*}

\begin{align*}
i&\mathcal{M}_s(g^a g^b \to g^c g^d)=(\epsilon_{d,4}^*)_\sigma[-g f^{edc}C^{\tau'\sigma \nu}(q_4-q_3,q_4,q_3)](\epsilon_{c,3}^*)_\nu\cdot\\
&\cdot\left[-i\frac{g_{\tau \tau'}-(q_\tau^sq_{\tau'}^s)/M_g^2}{(q_1+q_2)^2-M_g^2}\right](\epsilon_{b,2})_\mu [-g f^{eba}C^{\tau\mu\lambda}(q_2+q_1,-q_2,-q_1)](\epsilon_{a,1})_\lambda,
\end{align*}

\begin{align*}
i&\mathcal{M}_u(g^a g^b \to g^c g^d)=(\epsilon_{d,4}^*)_\sigma[-g f^{edb}C^{\tau'\sigma \mu}(-q_4+q_2,q_4,-q_2)](\epsilon_{b,2})_\mu\cdot\\
&\cdot\left[-i\frac{g_{\tau \tau'}-(q_\tau^uq_{\tau'}^u)/M_g^2}{(q_2-q_4)^2-M_g^2}\right](\epsilon_{c,3}^*)_\nu [-g f^{eac}C^{\tau\lambda\nu}(q_1-q_3,-q_1,q_3)](\epsilon_{a,1})_\lambda,
\end{align*}

\begin{align*}
i\mathcal{M}_4&(g^a g^b \to g^c g^d)=\\
&-ig^2[f^{abe}f^{cde}(g^{\lambda \nu}g^{\mu \sigma}-g^{\lambda \sigma}g^{\mu\nu})+f^{ace}f^{bde}(g^{\lambda \mu}g^{\nu \sigma}-g^{\lambda \sigma}g^{\mu\nu})-\\
&f^{ade}f^{cbe}(g^{\lambda \mu}g^{\sigma \nu}-g^{\lambda \nu}g^{\sigma\mu})](\epsilon_{d,4}^*)_\sigma(\epsilon_{c,3}^*)_\nu(\epsilon_{b,2})_\mu(\epsilon_{a,1})_\lambda.
\end{align*}

The invariant matrix element, after proper averaging over the initial (summing over the final) gluon states is:
$$|\mathcal{M}(gg\to gg)|^2=\frac{1}{8\times 2}\frac{1}{8\times 2}\sum_{\text{colour}}\sum_{\text{pol.}}|\mathcal{M}_t+\mathcal{M}_u
+\mathcal{M}_s+\mathcal{M}_4|^2.$$

The final result is:

\begin{align*}
|\mathcal{M}(&gg\to gg)|^2=\\
   9g^4&[56160M_g^{16}-100224M_g^{14}(t+u)+\\
   48M_g^{12}&(2111t^2+1934tu+2111u^2)-\\
1208M_g^{10}&(49t^3+53t^2u+53tu^2+49u^3)+\\
M_g^8&(20559t^4+30182t^3u+34705t^2u^2+30182tu^3+20559u^4)-\\
M_g^6&(3948t^5+8743t^4u+12266t^3u^2+12266t^2u^3+8743tu^4+3948u^5)+\\
M_g^4&(329t^6+1059t^5u+2035t^4u^2+2289t^3u^3+\\
&2035t^2u^4+1059tu^5+329u^6)-\\
2M_g^2&tu(3t^5+15t^4u+29t^3u^2+29t^2u^3+15tu^4+3u^5)+\\
&t^2u^2(t+u)^2(t^2+tu+u^2)]/[512M_g^4(M_g^2-s)^2(M_g^2-t)^2(M_g^2-u)^2].
\end{align*}

\addcontentsline{toc}{chapter}{\iflanguage{italian}{Bibliografia}{Bibliography}}
% class option: science/humanities toggles ssc-num/ssc-alpha for the bibliographystyle
\bibliography{biblio}

%\chapter*{\iflanguage{italian}{Ringraziamenti}{Acknowledgements}}
%\addcontentsline{toc}{chapter}{\iflanguage{italian}{Ringraziamenti}{Acknowledgements}}

%\lipsum[20]

\newpage

\begin{textblock*}{176mm}(0pt,0pt)%
\noindent\includegraphics*[height=250mm]{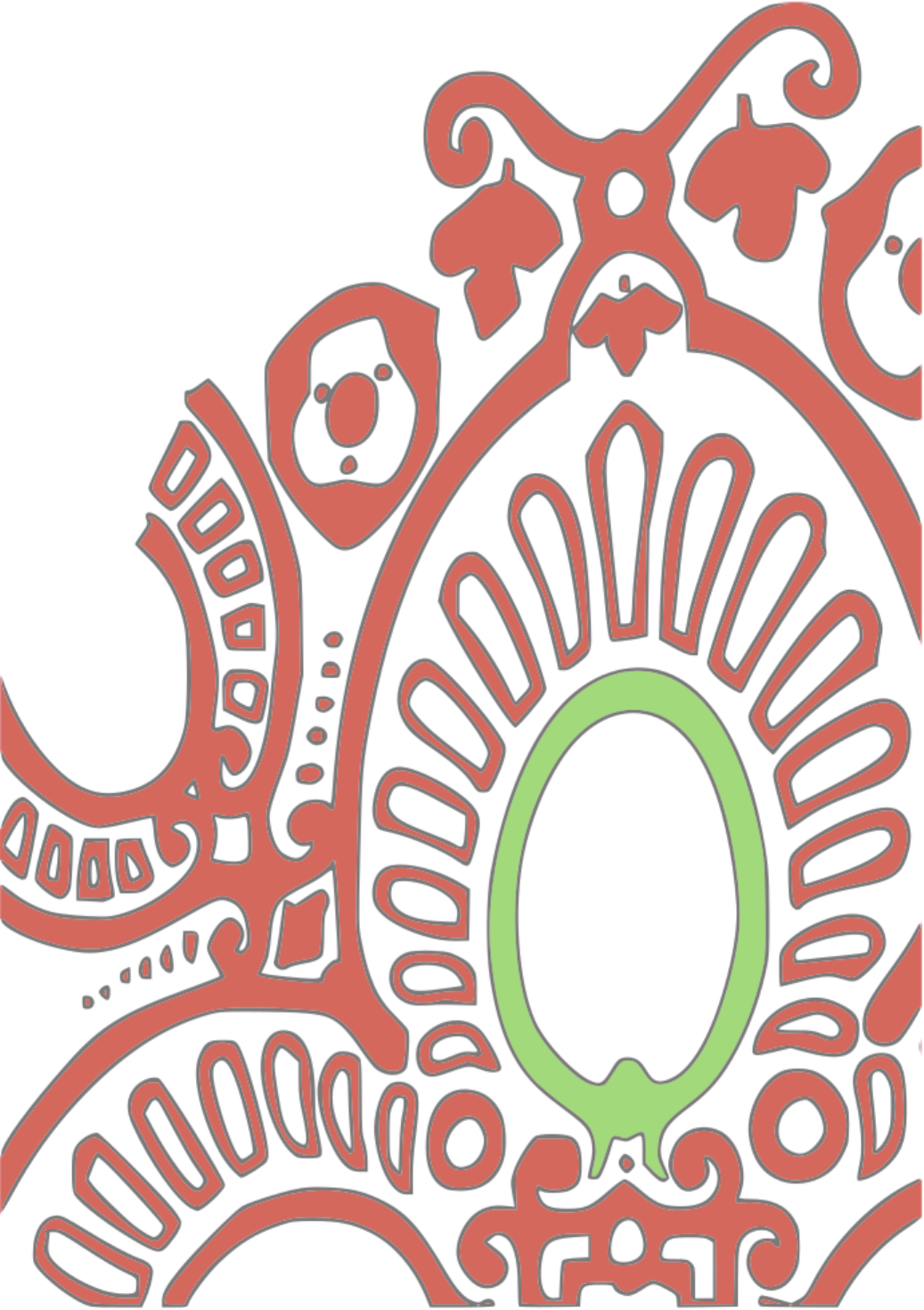}
\end{textblock*}

% comment out this line, if you don't want the last "cover" page
\end{document}